\documentclass[aps,prx,twocolumn,superscriptaddress,longbibliography,nofootinbib]{revtex4-2}
\usepackage{graphicx,amssymb,amsmath,amsfonts,empheq,latexsym,braket,bm,bbm,dsfont,upgreek}
\usepackage[normalem]{ulem}
\usepackage[dvipsnames]{xcolor}
\usepackage{comment}
\usepackage{tabularx}
\usepackage{color}
\usepackage[dvipsnames]{xcolor}
\usepackage{makecell}
\usepackage{multirow}
\usepackage{url}
\usepackage{soul}
\usepackage{ulem}
\usepackage{algorithm}
\usepackage{enumitem}
\usepackage{algpseudocode}
\usepackage{hyperref}
\hypersetup{
colorlinks = true,
linkcolor = [rgb]{0,0,0}, 
citecolor = [rgb]{0.13,0.55,0.13},
urlcolor = [rgb]{0.25, 0.41, 0.88}}

\usepackage{tikz}
\newcommand{\rrangle}{\rangle\!\rangle}
\newcommand{\llangle}{\langle\!\langle}

\newcommand{\tr}{\operatorname{tr}}

\renewcommand{\tilde}{\widetilde}

\def\tr{\text{tr}}
 
\renewcommand\[{\begin{equation}}
\renewcommand\]{\end{equation}}

\newcommand{\bbI}{\mathbb{I}}

\newcommand{\bbZ}{\mathbb{Z}}

\newcommand{\calD}{\mathcal{D}}

\newcommand{\calN}{\mathcal{N}}
\newcommand{\calO}{\mathcal{O}}

\newcommand{\eqnref}[1]{Eq.\,\eqref{#1}}
\newcommand{\figref}[1]{Fig.\,\ref{#1}}

\newcommand{\secref}[1]{Sec.\,\ref{#1}}
\newcommand{\appref}[1]{Appendix.\,\ref{#1}}

\begin{document}

\title{Fractional quantum Hall states under density decoherence}

\author{Zijian Wang}
\thanks{ZW and RF contributed equally to this work.}
\affiliation{Institute for Advanced Study, Tsinghua University, Beijing, China}
\affiliation{Department of Physics, University of California, Berkeley, CA 94720, USA}
\author{Ruihua Fan}
\thanks{ZW and RF contributed equally to this work.}
\affiliation{Department of Physics, University of California, Berkeley, CA 94720, USA}
\author{Tianle Wang}
\affiliation{Department of Physics, University of California, Berkeley, CA 94720, USA}
\affiliation{Department of Physics, Harvard University, Cambridge, MA 02138, USA}
\author{Samuel J. Garratt}
\affiliation{Department of Physics, University of California, Berkeley, CA 94720, USA}
\affiliation{Department of Physics, Princeton University, NJ 08544, USA}
\author{Ehud Altman}
\affiliation{Department of Physics, University of California, Berkeley, CA 94720, USA}
\affiliation{Materials Sciences Division, Lawrence Berkeley National Laboratory, Berkeley, CA 94720, USA}

\begin{abstract}  

Fractional quantum Hall states are promising platforms for topological quantum computation due to their capacity to encode quantum information in topologically degenerate ground states and in the fusion space of non-abelian anyons. 
We investigate how the information encoded in two paradigmatic states, the Laughlin  and Moore-Read states, is affected by density decoherence -- coupling of local charge density to non-thermal noise. 
We identify a critical filling factor $\nu_c$, above which the quantum information remains fully recoverable for arbitrarily strong decoherence. The $\nu=1/3$ Laughlin state and $\nu = 1/2$ Moore-Read state both lie within this range.
Below $\nu_c$ both classes of states undergo a decoherence induced Berezinskii-Kosterlitz-Thousless (BKT) transition into a critical decohered phase.
For Laughlin states, information encoded in the topological ground state manifold degrades continuously with decoherence strength inside this critical phase, vanishing only in the limit of infinite decoherence strength.
On the other hand, quantum information encoded in the fusion space of non-abelian anyons of the Moore-Read states remains fully recoverable for arbitrary strong decoherence even beyond the BKT transition. 
These results lend further support to the promise of non-Abelian FQH states as platforms for topological quantum computation and raises the question of how errors in such states can be corrected.
\end{abstract}

\maketitle
\tableofcontents
\newpage
\section{Introduction}   

One route towards realizing topological quantum computation is through solid-state systems that host anyonic excitations~\cite{Kitaev:2000nmw,Kitaev:2005hzj,ChetanReview,Alicea:2012review,Karzig:2016oji,Alicea:2024}. A paradigmatic example is the fractional quantum Hall (FQH) platform, where charged anyons can be manipulated with electrostatic controls to perform quantum computation~\cite{prange1989quantum,halperin2020fractional,DasSarma:2005zz,Walker:2005,Bravyi:2005}.
There are two major sources of errors that arise in such devices, one is from thermally excited quasiparticles, and the other from decoherence caused by local noise in the device, cosmic rays, and other uncontrolled effects.
The thermal errors can be suppressed by cooling the system and safely ignored at experimentally accessible low temperatures~\cite{Alex:2023gap,Yazdani:2023gap}. 
However, the second type of error is harder to eliminate and persists throughout the experiment.
In this work, we examine the effect of such local decoherence on the encoded quantum information.

There are two ways to encode quantum information in a topologically ordered FQH state~\cite{ChetanReview}.
The first is to store the information in the topologically degenerate ground state subspace, which applies to both Abelian and non-Abelian FQH states.
The second is to prepare a non-Abelian quantum Hall state and encode information into the fusion space of non-Abelian anyons.
In practice, the first route may require using bilayer FQH systems or fractional quantum spin Hall insulators with engineered edge couplings~\cite{lindner2012fractionalizing,barkeshli2016charge,wen2024cheshire}. 
For the purpose of analyzing robustness against decoherence, it is more convenient to consider a single-layer FQH states on a torus as a theoretical simplification.  
We will investigate the robustness of quantum information stored in both approaches.

From a theoretical perspective, our investigation aligns with the broader program of understanding topological order in mixed quantum states~\cite{fan2023diagnostics,bao2023mixed, lee2023quantum, wang2025intrinsic, sohal2025noisy, ellison2025towards, sang2024mixed, sang2025stability, sang2025mixed, yang2025topological, zhang2025strong, lessa2025higher, chen2024unconventional,sala2025decoherence, sala2025stability,katsura2020fate,li2025howmuch,tang2025phases}.
Quantifying the capacity for encoding quantum information amounts to characterizing the long-range entanglement in the presence of decoherence.
We can then view it as a probe of topological order in the mixed-state setting. Most prior work in this direction has focused on fixed-point models with commuting projector Hamiltonians.
As we will see below,  decohered FQH states are different in several important ways, raising new technical and conceptual challenges. 
Moreover, decoherence-induced transitions of non-Abelian topological orders remain much less explored than the corresponding transitions in Abelian states~\cite{sala2025decoherence, sala2025stability}. In particular, little is known about the fate of non-Abelian anyons under decoherence. This is one of the central questions addressed in this work.

For topological stabilizer codes or the quantum double models, one can always specify a decoding algorithm and discuss its success probability~\cite{Dennis:2001nw,Gottesman2009,TerhalReview,Fujii:2015wia}. 
While the non-stabilizer and chiral nature of FQH states complicates an explicit construction of this kind, we can infer the threshold of the optimal decoder by analyzing certain information-theoretic quantities~\cite{fan2023diagnostics}. In particular, we can use the quantum coherent information to quantify the recoverable quantum information in a decohered system~\cite{Schumacher:1996dy,Schumacher:2001}. 
A sharp drop in this quantity as the decoherence increases beyond a threshold signals the breakdown of the FQH state as a quantum memory~\cite{fan2023diagnostics}.

In realistic platforms such as graphene heterostructures, the decoherence in the bulk conserves the total electron number and can couple only to the local charge density or current.
Here, we focus on density dephasing, i.e., noise coupled to the microscopic particle density, and assume it is uniform in space. 
After the ground state is prepared, the subsequent dynamics have two components. 
First, decoherence continually creates small pairs of anyons. Second, these anyons move diffusively under the Hamiltonian evolution. Here, we isolate the effect of the first part by neglecting the Hamiltonian evolution entirely. Such a simplified dynamics is described by the following Lindblad equation
\begin{equation}
    \frac{d\hat\rho}{dt} = \gamma\int d^2 z \big( \hat{n}(z)\, \hat{\rho}\, \hat{n}(z)-\frac{1}{2}\{\hat n(z)^2,\hat \rho\} \big)\,,
    \label{eq:density dephasing}
\end{equation}
where $\hat\rho$ is the density matrix, $\hat{n}(z)$ the electron density operator. 
By dimensional analysis, the decoherence strength $\gamma$ has the unit of $[\text{time}]^{-1}[\text{length}]^2$, where the length scale can only come from the short-distance regularization. 
Therefore, it is important to keep a \emph{finite} short-distance cutoff to ensure that the density decoherence has non-vanishing effects.\footnote{It is natural to expect that the universal physics is insensitive to the precise form of the regularization. One specific and possibly physical choice is to use the lowest-Landau-Level projected density operator in \eqnref{eq:density dephasing} and the short-distance length scale is the magnetic length. We will comment on its effect at the end of the manuscript.}

Specifically, we examine the effect of decoherence in the Laughlin and Moore-Read states, the simplest Abelian and non-Abelian FQH states, respectively~\cite{laughlin1983,moore1991Nonabelions}. 
the Laughlin state offers the simplest arena for analyzing the robustness of information encoded in degenerate ground states and also help with the development of tools before addressing the more intricate non-abelian states. 
The Moore-Read state is the simplest non-abelian FQH state that can be used for quantum computation with anyons.
We derive our results by using the well-known model wave functions of these states. 
They admit well-developed plasma analogies and conformal field theory (CFT) descriptions that allow us to solve the problem analytically~\cite{laughlin1983,moore1991Nonabelions,Read:2008rn,bonderson2011Plasma}.
One can also employ the Chern-Simons theory description of the quantum Hall states and derive the same result. Therefore, we expect our result to be applicable even beyond the ideal wave functions.

The rest of the paper is organized as follows. In Sec.~\ref{sec:overview} we provide a non-technical overview of the main results and motivate them with physical arguments, leaving the technical derivations to subsequent sections. In Sec.~\ref{sec:Laughlin} we use the plasma analogy to identify and characterize a decoherence-induced transition in Laughlin states. Then, in Sec.~\ref{sec:laughlin_memory} we determine how the quantum coherent information, quantifying the recoverable quantum memory, evolves across the transition. To accomplish this, we utilize the field theory description of the Laughlin state in terms of a 1+1-dimensional CFT. In Sec.~\ref{sec:Moore Read} we use the CFT formulation to identify the decoherence-induced transition in Moore-Read states and determine their effect on information encoded in the fusion space of non-abelian anyons. We conclude with general remarks on the results and outlook in Sec.~\ref{sec:discussions}.

\section{Overview}
\label{sec:overview}

In this section, we give an overview of the main results, emphasizing the physical intuition.
Our starting point is an FQH state that encodes quantum information either in the topological ground state degeneracy or the fusion space of non-abelian anyons. We assume that this logical qudit is entangled in some way with another qudit $R$, as it would be if it were taking part in a quantum computation. 
The amount of information that can be recovered using operations on the system $Q$, subject to decoherence, can be quantified by the quantum coherent information $I_c(R\rangle Q)=S_Q-S_{RQ}$, where $S_Q$ and $S_{RQ}$ are the von Neumann entropy of these systems after the decoherence. 

The density matrix of the system affected by the decoherence channel \eqref{eq:density dephasing} has a simple form in the particle position basis:   
\begin{equation}
    \rho(z^1,z^2) =\Psi(z^1)\Psi^\star(z^2) e^{-\frac{1}{2}V(z^1,z^2)} \,.
    \label{eq:rho}
\end{equation}
Here $z^{\alpha}\equiv\{z^\alpha_1,\ldots ,z^\alpha_N\}$ denote electron configurations, $\Psi(z^\alpha)$ is the first quantized wavefunction of the original pure state, and the decoherence generates $V$ that depends on the electron density $n_i(z)$ as
\begin{equation}
    V(z^1,z^2) = \gamma t \int d^2 z (n_1(z) - n_2(z))^2\,.
    \label{eq:V attraction}
\end{equation}
It is convenient to define a dimensionless measure of the decoherence strength $\mu=\gamma t /2\pi a^2$, where $a$ is a short distance cutoff much smaller than the magnetic length. In the strong decoherence limit $\mu\rightarrow +\infty$, the coupling $V$ suppresses off-diagonal elements and thus the density matrix reduces to a classical probability distribution. 

\begin{figure}
    \centering
    \includegraphics[width=\linewidth]{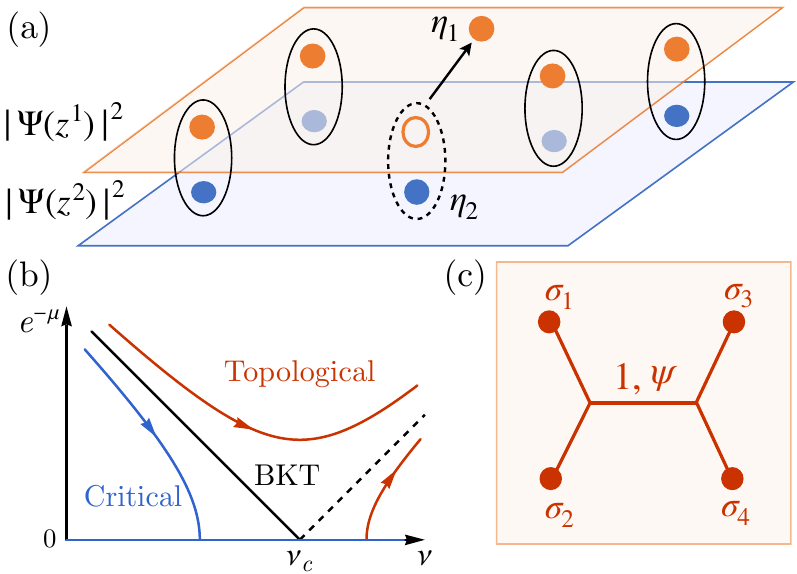} 
    \caption{(a) Schematics for the purity of the decohered state. Two layers are the two copies of states, orange and blue dots are the electrons, black circles represent their tight binding at the strong decoherence limit, dashed circles is the unbinding event when deviating from the limit. (b) Renormalization-group flow for decohered Laughlin and Moore-Read states. Below a critical filling the system enters a critical phase through a BKT transition. Above it the system remains topological. (c) Encoding a single qubit in the two-dimensional fusion space of four well-separated fundamental quasiholes of the Moore-Read state.}
    \label{fig:intro fig 1}
\end{figure}

The simplest proxy for the behavior of the quantum coherent information is the R\'enyi-2 version of this quantity. Thus, as a first step, we analyze the purity as a generating function for all other R\'enyi-2 information quantities. Following Eq.~\eqref{eq:rho} we have
\begin{equation}
    \text{tr}\rho^2= \int \calD z^1\calD z^2|\Psi(z^1)|^2|\Psi(z^2)|^2 e^{-V(z^1,z^2)},
    \label{eq:purity}
\end{equation}
which can be interpreted as a partition function of {\it classical} electron fluids residing in two layers as depicted in \figref{fig:intro fig 1}~(a). The intra-layer interactions among fluid particles derive from the structure of the initial state $\Psi(z)$ through the modulus squared. At the same time, the decoherence introduces an inter-layer attraction \eqref{eq:V attraction}, which grows with the decoherence strength.  
Thus, in the pure initial state, electrons in the two layers fluctuate independently, while in the opposite limit $\mu\rightarrow +\infty$ they form tightly bound pairs, depicted as the solid circles in \figref{fig:intro fig 1}~(a).
The nature of the transition between the two extremes can be understood by expanding about the strong decoherence limit in the small fugacity $e^{-\mu}$ of unbound interlayer pairs.

\subsubsection{Decohered Laughlin States}

The physics of unlocking the two electron fluids near the strong decoherence limit is particularly intuitive in the Laughlin states with filling factor $\nu=1/m$. This is addressed in Sec.~\ref{sec:Laughlin}.  
Through Laughlin's plasma analogy, the wavefunction gives rise to Coulomb interactions between particles of effective charge $\sqrt{2m}$ within each layer. In the pure state $\mu=0$, the two plasmas are independently in their screening phase (for $m<70$), while in the opposite limit $\mu\to \infty$ they are locked into tightly bound pairs, which screen only the symmetric component of the charge. 
Because the interaction between unbound (relative) charges is not fully screened in the decohered state, their tendency to proliferate depends on the usual competition between the Coulomb energy cost and entropy gain, leading to the Berezinski-Kosterlitz-Thouless (BKT) RG flow depicted in \figref{fig:intro fig 1}(b)~\cite{berezinskii1972destruction,Kosterlitz1973,Kosterlitz1974}. Interestingly, the filling factor $\nu$ plays the role of the dielectric constant or inverse stiffness. So, increasing the decoherence strength moves the physical system (i.e., the initial condition for the flow) vertically downward on this diagram. This results in a BKT transition at a finite value of $\mu$ if $\nu<\nu_c$. Remarkably, if the filling factor $\nu\geq \nu_c$ the screening phase persists to arbitrarily large $\mu$ ($\nu_c=1/4$ in the analysis of the purity). This is suggestive, and we will show that it indeed implies, that a fully recoverable quantum memory persisting to arbitrarily strong decoherence. 

An important aspect of our analysis is that the fully decohered state falls on a  line of critical states, with properties determined by universal aspects of the FQH state. In fact, the expansion around the fully decohered state, which reveals the decoherence-induced transitions, relies on the same universal properties rather than on the detailed structure of the underlying state. Therefore, the main results of this analysis are expected to hold for FQH states in the Laughlin series beyond specific model wavefunctions.

The critical decohered state, discussed in Sec.~\ref{sec:critical}, is characterized by algebraic decay of R\'enyi correlations in both the electron and quasi-hole (anyon) operators, which originate from universal properties of the initial pure state.
It is interesting to connect this result with the understanding that decoherence-induced transitions  arise from anyon condensation in 2+0 dimensions \cite{bao2023mixed}. In the decohered Laughlin state anyons cannot fully condense because they carry a (fractional) charge of the global charge $U(1)$ symmetry. Instead, they form a critical condensate exhibiting quasi-long-range order.

The critical nature of the fully decohered state also helps us generalize from a theory of the purity to one of R\'enyi-1 and von Neumann quantities, which have more transparent information-theoretic meanings. This is done in Sec.~\ref{sec:laughlin_vN}. 
The key, again, is to develop an expansion in small fugacity about the diagonal, fully decohered state. Divergence of the expansion then signals an instability of the critical diagonal density matrix toward the encoding phase. While we are not able to obtain the exact critical filling factor, we derive rigorous bounds placing it within the range $1/8\leq \nu_c\leq 1/4$. 

So far we have not addressed directly how these transitions affect the quantum information encoded in the $m$-dimensional ground state subspace on a torus. To access these $m$ states, which are not easily discerned within the plasma analogy,  we reformulate the problem in Sec.~\ref{sec:laughlin_memory} in terms of an effective field theory, using the well known correspondence between the FQH states and correlations in 1+1 dimensional CFTs. In particular, a basis of $m$ Laughlin states on a torus is given by 
\begin{equation}
    \Psi_\ell(z)=\langle \prod_{i=1}^{N} e^{im\varphi(z_i) }e^{-\frac{i}{2\pi}\int d^2 z\varphi}\rangle_\ell,
\end{equation}
where the $N$-point correlation is evaluated in the CFT of a  chiral boson $\varphi$ with a twist of $2\pi\ell/m$ in the boundary conditions along $\hat{x}$ and $\ell=0,\ldots,m-1$.

In the expression for the purity \eqref{eq:purity}, the chiral and anti-chiral modes originating from the two wavefunctions and their conjugates add up to two non-chiral bosonic modes $\phi_\pm$. These fields can be interpreted as the phases conjugate to the plasmas associated with the symmetric and anti-symmetric (relative) charge between the two layers. 
The resulting field theory naturally reproduces the decoherence-induced transitions found through the plasma analogy. In the pure state ($\mu=0$), where the two plasmas are in the screening phase, both fields are locked to zero and the effective theory is gapped. In the opposite limit ($\mu\rightarrow\infty$), corresponding to the fully decohered state, only the symmetric charge density is screened in the plasma analogy, hence $\phi_+$ is still gapped, but $\phi_-$ is critical. 
Away from the infinite decoherence limit, the expansion of the purity in small fugacity leads to the low-energy effective sine-Gordon theory 
\begin{equation}
    Z_-=\int \calD \phi_- \exp\left(-\int \frac{m}{4\pi}(\nabla\phi_-)^2 - 2e^{-\mu}\cos m\phi_-\right)\,,
\label{eq:SG intro}
\end{equation}
which exhibits the same BKT transition discussed above~\cite{Jose1977}.  
The field theory formulation now allows us to compute the evolution of the coherent information across the transition separating the gapped screening phase, in which the cosine term is relevant, and the critical decohered state, in which it is irrelevant.

To have a non-trivial coherent information, the decoherence channel needs to operate on a Laughlin state on a torus $Q$, whose ground state manifold is entangled with a reference qudit $R$. In Sec.~\ref{sec:laughlin_memory} we show that the purities that enter the R\'enyi-2 coherent information can be written as sums over the sine-Gordon partition functions \eqref{eq:SG intro} with different twisted boundary conditions of the field $\phi_-$ or of its dual $\theta_-/m$. The R\'enyi-2 coherent information is then given by
\begin{equation}
I_c^{(2)}(R\rangle Q)=\log \frac{1+\sum_{\ell=1}^{m-1} e^{-\Delta F_{(0,\ell)}}}{1+\sum_{\ell=1}^{m-1}e^{-\Delta F_{(\ell,0)}}}\,.
\label{eq:Ic2 overview}
\end{equation}
Here $\Delta F(\ell,0)$ and $\Delta F(0,\ell)$ are the relative free energy costs of inserting a twist of $2\pi\ell/m$ in the boundary conditions of, respectively, $\phi_-$ and its dual $\theta_-/m$ along the $\hat{x}$ direction of the torus.

\begin{figure}
    \centering
    \includegraphics[width=\linewidth]{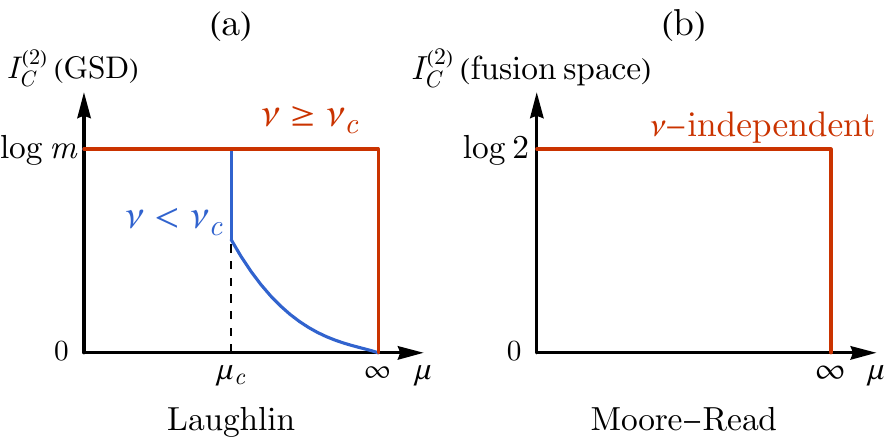}
    \caption{R\'enyi-2 quantum coherent information as a function of the decoherence strength. (a) For Laughlin states, the information is encoded in the degenerate ground states. For systems below the critical filling it starts to decrease after a finite time and becomes zero only at the strong decoherence limit. (b) For Moore-Read states, the information is encoded in the fusion space and is infinitely robust against the decoherence.}
    \label{fig:intro fig 2}
\end{figure}

In the screening phase ($\mu<\mu_c$) the cosine term dominates and any non-vanishing twist of $\phi_-$ incurs a free energy cost of order $L$, while twists in $\theta_-$ are free. Thus we have $I^{(2)}_c=\log m$ as expected. For $\mu>\mu_c$ the cosine term is irrelevant and the system flows to the critical phase. This however does not imply the vanishing of the coherent information. The contribution of the irrelevant cosine term to the free energy cost of a $\phi_-$ twist can be calculated perturbatively and is easily seen to be of order one. Thus, the coherent information jumps upon crossing the transition at $\mu_c$ from $\log m$ to a smaller non-universal positive value. As $\mu$ exceeds $\mu_c$, the coherent information decreases monotonously, as shown by the blue curve in Fig.~\ref{fig:intro fig 2}~(a), approaching zero in the limit $\mu\to\infty$. This result, a signature of the critical nature of the mixed state above the error threshold, stands in sharp contrast to the vanishing of the coherent information across the error threshold in the toric code \cite{fan2023diagnostics}.  

If the filling factor is above the critical filling $\nu_c$, the cosine term is relevant even at the infinite decoherence limit. Consequently, the coherent information stays at $\log m$ for any finite decoherence strength in the thermodynamic limit, as is shown by the red curve in Fig.~\ref{fig:intro fig 2}~(a). In a finite system of linear size $L$, however, the cosine term becomes effectively irrelevant when $\mu \sim \calO(\log L)$. As a result, the lifetime of the memory diverges logarithmically in the system size.

\subsubsection{Decohered Moore-Read states}

Moore-Read states are the simplest FQH states that host non-abelian anyons. Though these anyons do not support a fault-tolerant universal gate set, one can store protected quantum information in their fusion space, as is illustrated in Fig.~\ref{fig:intro fig 1}~(c). Our aim is to understand the decoherence-induced transitions in Moore-Read states and how they affect the information encoded in the fusion channels of non-abelian anyons. To this end, we use both an effective field theory description and a plasma analogy~\cite{bonderson2011Plasma} of Moore-Read states.

Like the Laughlin states, Moore-Read states can also be represented as a correlation function in a 1+1-dimensional CFT, which in this case includes two sectors: a chiral Ising (fermion) CFT and a chiral boson CFT that we refer to as the bosonic charge sector of the theory. 
For the description of the purity \eqref{eq:purity}, the chiral and anti-chiral modes originating from the wavefunctions and their conjugates, combine to a pair of {\it non-chiral} Ising CFTs and a pair of {\it non-chiral} boson CFTs. The purity is then formulated as an insertion of strong perturbations in this CFT. 

Our strategy again is to expand about the limit $\mu\to\infty$ of the fully decohered Moore-Read state. In this limit, the non-chiral boson field $\phi_-$ from the bosonic charge sector is in a free critical state, while the symmetric boson field $\phi_+$ is gapped, exactly as in the fully decohered Laughlin state. The Ising sector, on the other hand, forms a more complex strongly coupled theory. 
Nonetheless, we can obtain the long-range correlations in this sector, needed for the small fugacity expansion, by utilizing the plasma analogy of the Moore-Read states~\cite{bonderson2011Plasma}. 

Having established the correlations in the fully decohered state we use them to formulate the effective theory through the expansion about this reference point. In the bosonic charge sector, this gives rise to an effective sine-Gordon theory similar to \eqref{eq:SG}, which leads to a decoherence-induced BKT transition for filling factors $\nu<\nu_c$. 
In the Ising sector, however, the expansion about the fully decohered state inevitably generates relevant perturbations. As a result, for any finite decoherence the Ising sector flows back to a topological phase. This suggests the establishment of a new intrinsically mixed phase of matter, in which quantum information is preserved only in the non-abelian sector.

In Sec.~\ref{sec:MR_anyon} we verify the robustness of the quantum memory encoded in the fusion space of (four) non-abelian anyons by directly calculating the quantum coherent information. As a first step, we represent the basis states in this space as conformal blocks in the chiral Ising$\times U(1)$ CFT. This allows us to entangle the initial state with a reference qubit and represent the quantum coherent information of the decohered system in terms of expectation values in the effective field theory. The result, illustrated in Fig.~\ref{fig:intro fig 2}~(b), shows that the coherent information remains $\log 2$ for arbitrary (yet finite) decoherence strength, indicating fully recoverable quantum memory independent of the filling factor. 

At the same time, quantum information encoded in the bosonic charge sector of the degenerate ground state on a torus undergoes a transition similar to that found in the Laughlin state if $\nu<\nu_c$. Using the expansion in small fugacity about the von Neumann entropy of the fully decohered state we find that this critical filling for the Moore-Read states is bounded by $1/8\le \nu_c \le 1/3$.

\section{Decohered Laughlin states}
\label{sec:Laughlin}

In this section, we apply the plasma analogy to study decoherence-induced phase transitions in Laughlin states and show how they depend on the filling factor. We start with analyzing the purity of the density matrix, the R\'enyi-2 quantity, and then the von Neumann quantities.

\subsection{Plasma analogy and filling-dependent transition}
\label{sec:laughlin plasma analogy}

The Laughlin state is an incompressible state in the lowest landau level at filling $\nu=1/m$. Its universal properties are captured by the following un-normalized model wave function
\begin{equation}
\Psi(\{z_i\})=\prod_{i<j}^N (z_i-z_j)^m e^{-\sum_i|z_i|^2/4\ell_B^2}\,.
\end{equation}
Here, $m$ is an odd integer , $N$ is the number of electrons, $z_i=x_i+iy_i$ is the electron coordinate and $\ell_B$ is the magnetic length. Notably, all analysis in this work also apply to bosonic Laughlin states, where $m$ is an even integer.
To simplify the notations, we often omit the curly bracket $\{\cdots\}$ and set $\ell_B = 1$ from now on.
The norm square of $\Psi$ takes the form of a partition function of a single-component two-dimensional plasma
\begin{equation}
\begin{gathered}
\langle\Psi|\Psi\rangle = \int \calD z\, e^{-\Phi_L(z_i, \sqrt{2m})}\,,\\
\Phi_L(z_i, Q) = -Q^2\sum_{i<j}\log |z_i-z_j| + \frac{Q^2}{4m}\sum_i|z_i|^2\,,
\end{gathered}
\label{eq:laughlin plasma}
\end{equation}
where $\calD z \equiv \prod_{i=1}^N d^2 z_i/N!$ is the integration measure. In this analogy, each particle carries a charge $Q=\sqrt{2m}$ and moves on a uniform neutralizing background of charge density $-Q / 2m\pi$. 
For physically relevant values of $m$ (e.g., $m < 70$), the plasma is in a screening phase~\cite{caillol1982MonteCarloStudy}, and all long-distance properties of the Laughlin state follow. 

Following Eq. \eqref{eq:purity}, the purity admits a similar mapping to the partition function of a classical two component plasma, from which many properties of the decohered systems follow. 
Specifically, at time $t$ subject to the  decoherence \eqref{eq:density dephasing}, the purity of the Laughlin state is given by 
\begin{equation}
    \tr\hat{\rho}^2 =\int \calD (z^1,z^2) e^{- \Phi_L(z^1_i, \sqrt{2m}) - \Phi_L(z^2_i, \sqrt{2m}) - V}\,,
\label{eq:purity_Laughlin}
\end{equation}
where $\Phi_L(z^1_i, \sqrt{2m})$ and $\Phi_L(z^2_i, \sqrt{2m})$ are the plasma energy functionals defined in \eqnref{eq:laughlin plasma}.
It is therefore natural to interpret the purity as the partition function of two layers of plasmas, labeled by 1 and 2, which originate from the bra and ket space of the density matrix. 
The decoherence suppresses off-diagonal elements of the density matrix in the position basis.
In the language of plasma, it induces an attractive interaction between the two plasmas, through the potential 
\begin{equation}
    V = \gamma t \int d^2 z (n_1(z) - n_2(z))^2\,.
\end{equation}
As noted in the overview, $V$ depends on the short-distance regularization. 
One convenient choice is to smear the density operator in \eqnref{eq:density dephasing} by a Gaussian
\begin{equation}
    \hat{n}(z) \mapsto \frac{1}{\pi a^2} \int d^2 z' e^{-|z-z'|^2/a^2} \hat{n}(z')\,,
    \label{eq:regularization}
\end{equation}
which yields
\begin{equation}
V = \mu \sum_{i,j=1}^N e^{-\frac{|z^1_i-z^1_j|^2}{2a^2}} + e^{-\frac{|z^2_i-z^2_j|^2}{2a^2}} - 2 e^{-\frac{|z^1_i-z^2_j|^2}{2a^2}}\,,
\label{eq:fugacity}
\end{equation}
where $\mu = \gamma t / 2\pi a^2$ is a the dimensionless quantity that parameterizes the decoherence strength.
The first two terms in \eqnref{eq:fugacity} describe a short-range repulsion within each plasma, which is overwhelmed by the Coulomb repulsion and does not qualitatively affect the physics. 
Below we examine how the last term, representing interlayer attraction, affects the bilayer plasma.

An instructive reference point for understanding the transition is the the strong decoherence limit ($\mu \rightarrow +\infty$). 
In the plasma analogy, this limit corresponds to having infinitely strong inter-layer attractions so that all charges in the two layers form tightly bound pairs. 
Denoting the density matrix in this limit by $\rho_{\infty}$, we have
\begin{equation}
    Z_{\infty} = \tr \rho_\infty^2 = \int \calD z e^{-2\Phi_L(z_i, \sqrt{2m})}\,,\quad \mu \rightarrow +\infty\,,
    \label{eq:laughlin strong decoherence limit}
\end{equation}
where $z_i = z^1_i=z^2_i$ labels the position of each pair.\footnote{\eqnref{eq:laughlin strong decoherence limit} implicitly depends on the short-distance cutoff $a$. First, it holds only when the distances between these pairs are larger than $a$. Second, we omit a prefactor $(\kappa a^2)^{N}$ that accounts for the residual relative motion of each pair. Namely, the two particles in each pair can still move relative to each other within a disk of radius $a$. And $\kappa$ is a non-universal geometric factor.}
We can re-interpret $Z_\infty$ as the partition function of a single-component plasma formed by the pairs, i.e. $2\Phi_L(z_i, \sqrt{2m})=\Phi_L(z_i,2\sqrt{m})$. Each particle carries a charge $Q=2\sqrt{m}$ and moves on a neutralizing background.
For $m < 35$, any test charge that is only coupled to this paired plasma is fully screened, while test charges coupled to one of the two plasmas are partially screened. 
As a result, the plasma is in a different phase than at $\mu = 0$.

As a first step to characterizing the behavior at finite $\mu$, where unbound pairs may exist, we examine the screening of external test charges in the $Z_\infty$ plasma. 
Specifically, consider inserting two pairs of test charges: $q_1$ and $q_2$ at position $\eta$ in the two respective layers, and $-q_1$ and $-q_2$ at $\eta'$. The coupling to the plasma mediates an effective interaction between the test charges:
\begin{equation}
\begin{aligned}
    &e^{-U_{\text{eff}}(\eta,\eta')}\\
    =& \frac{1}{Z_{\infty}} \int \calD z   \exp\{-\Phi_L(z_i, \sqrt{2m};\eta,q_1;\eta',-q_1)\\
    &\qquad\qquad\qquad~-\Phi_L(z_i,\sqrt{2m};\eta,q_2;\eta',-q_2)\} \,.\\
\label{eq:test_charge}
\end{aligned}
\end{equation}
Here $\Phi_L(z_i,Q; \eta,q;\eta',q')$ denotes the energy of a charge-$Q$ single-component plasma with two test charges $q,q'$ inserted at $\eta,\eta'$:
\[
\begin{aligned}
\Phi_L(z_i,Q; & \eta,q;\eta',q') \\
=& \Phi_L(z_i, Q) +\frac{Qq}{4m}|\eta|^2+\frac{Qq'}{4m}|\eta'|^2 \\
& - \sum_i Qq\log|z_i- \eta| + Qq'\log|z_i-\eta'|\\
& - q q' \log |\eta - \eta'|\,,
\end{aligned}
\label{eq:laughlin_energy_testcharge}
\]
where the three lines of additional terms are the interaction energies with the background, mobile particles, and the bare interaction between test charges themselves.
We can formally separate the test charges into a pair of ``center-of-mass" charges $q_{\text{CM}} = \pm (q_1 + q_2)/\sqrt{2}$ that are coupled only to the charge-$2\sqrt{m}$ paired plasma, and a pair of ``relative" charges $q_{\text{rel}} = \pm (q_1 - q_2)/\sqrt{2}$ that interact only with each other. Namely, we rewrite the exponent in \eqnref{eq:test_charge} as
\[
\begin{aligned}
    -\Phi_L\big(z_i, 2\sqrt{m};  \eta,|q_{\text{CM}}| &;\eta',-|q_{\text{CM}}|\big) \\
    & -|q_{\text{rel}}|^2 \log|\eta-\eta'|\,,
\end{aligned}
\]
and accordingly
\begin{equation}
    U_{\text{eff}}(\eta,\eta') = \Delta F(\eta,\eta') + |q_{\text{rel}}|^2 \log |\eta-\eta'|\,.
\end{equation}
The first term is the free energy difference of the charge-$2\sqrt{m}$ plasma with and without the pair of the ``center-of-mass" charges.  
Since the charge-$2\sqrt{m}$ plasma is in the screening phase for $m<35$, we expect $\Delta F(\eta,\eta')$ to approach some finite constant when $|\eta-\eta'|$ is much larger than the screening length, and thus 
\begin{equation}
    U_{\text{eff}}(\eta,\eta') = |q_{\text{rel}}|^2 \log |\eta-\eta'| + \ldots\,,
\label{eq:Coul_eff}
\end{equation}
where ellipsis are constant corrections.
We conclude that the ``relative" charges interact via unscreened Coulomb interaction. 

The last result suggests that at finite decoherence strength $\mu$ the bilayer plasma may undergo a BKT transition through unbinding of the relative charges. 
Following the standard BKT analysis, we estimate the free energy cost $f_0$ of each individual unpaired particle. To this end, we consider a single unbinding pair, with the two particles separated at a distance of the order of the linear size $L$.
The unbinding introduces a pair of ``center-of-mass" charges that are fully screened by the charge-$2\sqrt{m}$ plasma, and a pair of ``relative" charge $q_{\text{rel}} = \pm \sqrt{m}$ whose interaction energy is given by \eqnref{eq:Coul_eff}. 
The total energy cost of the unbinding event is then $2\mu + m\log (L/a) + \ldots$, where the ellipsis represent the constant terms due to screening. 
The entropy gain is $\log (NA/a^2)$, where $A$ is the total area, $N \propto A$ the total particle number. Thus for $m< 4$ the free energy cost of unbinding is negative, indicating that the strongly decohered state is unstable towards the unbinding of inter-layer pairs for arbitrarily small deviation from infinite decoherence (i.e. for any non vanishing fugacity $\propto e^{-\mu}$). We will show below that this also holds for $m=4$.

For a more systematic treatment of the BKT transition we expand the partition function, representing the purity near the strong decoherence limit, as a sum over the number of unpaired particles.
In the dilute limit, the unpaired particles effectively introduce relative charges $\pm \sqrt{m}$ with the interaction \eqref{eq:Coul_eff}. We can approximate the partition function by
\begin{equation}
\begin{gathered}
    Z = Z_{\infty} Z_{\text{rel}} \\
    Z_{\text{rel}} \approx \sum_{k}\frac{e^{-2\bar \mu k}}{(k!)^2}\int \prod_i^{2k}d^2 \xi_i e^{\sum_{i<j}q_i q_j\log |\xi_i - \xi_j|}\,.
\end{gathered}
\label{eq:laughlin_near_infty1}
\end{equation}
Namely, the original bilayer plasma re-organizes into a charge-$2\sqrt{m}$ paired plasma $Z_\infty$ in its screening phase (for $m<35)$ and a plasma of ``relative" charges $Z_{\text{rel}}$ in the grand canonical ensemble, with $q_i=\pm \sqrt{m}$,  $\sum_iq_i=0$, and fugacity $e^{-\bar\mu} \propto e^{-\mu}$.\footnote{The fugacity comes from the unbinding energy as well as the the entropic effect analyzed above, i.e., $e^{-\bar\mu}\sim e^{-\mu}\times \sqrt{N/Aa^2}$. We also omitted a $O(1)$ prefactor that depends on the details of the screening due to the charge-$2\sqrt{m}$ plasma and the short-distance cutoff.}
The phase diagram is determined by the standard RG flow of the BKT transition as shown in Fig.~\ref{fig:intro fig 1}~(b) with $\nu = 1/m$ taking the role of the dimensionless temperature (or inverse stiffness). Increasing the decoherence strength tunes the starting point of the RG flow downward on a vertical line in this phase diagram.  
Crucially, the RG flow implies that for $m\leq 4$, the ``relative" plasma is always in the screening phase for arbitrarily small fugacity. For $m>4$ on the other hand, there is a BKT transition at a finite noise strength $\mu_c$, leading to an unscreened phase at $\mu>\mu_c$. 

\subsection{Critical phase above the threshold}\label{sec:critical}
 
In this section we examine the critical phase.  
As explained in \cite{fan2023diagnostics}, only quantities that are non-linear functions of the density matrix can show the establishment of a new phase beyond the decoherence threshold. Here we characterize the critical phase using the R\'enyi-2 correlation functions~\cite{fan2023diagnostics,bao2023mixed,lee2023quantum,ma2025topological,lessaSWSSB,sala2024spontaneous}. 
In particular the R\'enyi-2 correlation for electrons is given by
\begin{equation}
C_e^{(2)}(\eta,\eta')=\frac{\tr \big(c(\eta')\rho c^\dagger(\eta')c(\eta)\rho c^\dagger(\eta)\big)}{\tr\rho^2}\,.
\label{eq:C2e}
\end{equation}
Intuitively, $C^{(2)}_{e}$ quantifies the distinguishability between the state $\rho$ with a hole at $\eta$ and another state with a hole at $\eta'$. Below the threshold, we expect $C^{(2)}_{e}$ to decay exponentially with $|\eta - \eta'|$. 
Above the threshold, however, we will show that $C^{(2)}_{e}$ decays algebraically. 

We start with the infinite decoherence limit ($\mu=+\infty$), where the numerator of \eqnref{eq:C2e} is given by 
\begin{equation*}
\begin{aligned}
\tr & \big( c(\eta') \rho c^\dagger(\eta') c(\eta)\rho c^\dagger(\eta) \big) =\frac{1}{(N-1)!} \\
& \int \prod_{i=1}^{N-1} d^2 z_i e^{-\Phi_L(z_i, 2\sqrt{m};\eta,\sqrt{m};\eta',\sqrt{m})- m\log|\eta-\eta'|}\,.
\end{aligned}
\end{equation*}
Similar to the analysis of Eq.~\eqref{eq:test_charge}, the R\'enyi-2 correlation can be rewritten as 
\begin{equation}
C^{(2)}_e(\eta,\eta')=\frac{N}{A}e^{-\Delta F'(|\eta-\eta'|)-m\log|\eta-\eta'|},
\end{equation}
where $N/A$ is the average particle density, and $\Delta F'$ is  is the free energy difference between inserting a pair of test charges with $q=+\sqrt{m}$ at $\eta,\eta'$ in a Laughlin plasma with $Q=+2\sqrt{m}$, and that of inserting a single test charge $+2\sqrt{m}$. Because the plasma is in the screening phase (for $m<35$), $\Delta F'$ is independent of the separation $|\eta-\eta'|$ at large separations, and we have
\begin{equation}
C^{(2)}_e (\eta-\eta')\sim \frac{1}{|\eta-\eta'|^{m}},\quad (\mu=\infty).
\label{eq:Renyi2_1}
\end{equation}
For $\mu_c\leq \mu < \infty \quad (m>4)$, the Luttinger parameter is renormalized by the cosine term, so 
\begin{equation}
C^{(2)}_e (\eta-\eta')\sim \frac{1}{|\eta-\eta'|^{\alpha}},\quad 4\leq \alpha<m
\label{eq:Renyi2_2}
\end{equation}
where $\alpha=4$ at the critical transition $\mu=\mu_c$. 

The power-law correlations 
in the decohered state are inherited from the power-law correlation of composite bosons in Laughlin states~\cite{Girvin1987off,zhang1989Effective}. 
For the Laughlin state with filling factor $\nu = 1/m$, the composite bosons $b$ are defined by attaching $m$ flux quanta to each electron via a singular gauge transformation:
\begin{equation}
    b(z) = c(z) e^{-im\int d^2 z' n(z')\text{arg}(z'-z)}.
\end{equation}
The pure Laughlin state has an off-diagonal quasi-long range order for such composite boson 
$$
    \langle \Psi|b^\dagger(\eta')b(\eta)|\Psi\rangle\sim \frac{1}{|\eta-\eta'|^{m/2}}\, .
$$ 
For the decohered Laughlin state, the R\'enyi-2 correlation of composite bosons has a simple plasma analogy for arbitrary dephasing strength:
\begin{equation*}
\begin{gathered}
 \frac{\tr \big( b(\eta')\rho b^\dagger(\eta') b(\eta)\rho b^\dagger(\eta) \big)}{\tr\rho^2}=\frac{1}{|\eta-\eta'|^m}\left(\frac{1}{(N-1)!}\right)^2\\
 \frac{\int \prod_{i=1}^{N-1} d^2z^1_i d^2z^2_i e^{-V-\sum_{\kappa =1,2} \Phi_L(z^\kappa_i, \sqrt{2m};\eta,\sqrt{m/2};\eta',\sqrt{m/2})}}{\int \calD (z^1,z^2) e^{-V-\sum_{\kappa =1,2} \Phi_L(z^\kappa_i, \sqrt{2m})}}
 \end{gathered}
\end{equation*}  
where the test charges at $\eta,\eta'$ carry only the ``cener-of-mass" charges and will be screened at long distance. Therefore, the correlator decays algebraically for any dephasing strength  
\[
\frac{\tr \big( b(\eta')\rho b^\dagger(\eta') b(\eta)\rho b^\dagger(\eta) \big)}{\tr\rho^2}\sim \frac{1}{|\eta-\eta'|^m}\,.
\]
The composite boson correlation functions are related to the electron correlations via the non-local phase factors implementing the flux attachment. However, at infinite dephasing the phase factors from the two sides of the density matrix exactly cancel because the fully dephased density matrix is diagonal in the local occupation. Therefore, in this limit the R\'enyi-2 electron correlations are the same as the R\'enyi-2 composite boson correlations, which leads to Eq.~\eqref{eq:Renyi2_1}. Ref.~\cite{lu2023mixed} noted a similar effect, where the power-law composite boson correlations can be translated to local observables through measurement and feedback.
 
We can find the correlation functions of quasi-holes in the critical state by replacing the electron operators in \eqnref{eq:C2e} with quasi-hole operators. 
Let $\rho_\eta$ be the decohered states with a single quasi-hole at $\eta$
\[
\begin{gathered}
\rho_\eta=e^{\mathcal{L}t}[|\Psi_\eta\rangle\langle \Psi_\eta|], \\
\Psi_\eta(\{z_i\})=\prod_i^N(z_i-\eta)\prod_{i<j}^N (z_i-z_j)^m e^{-\sum_i|z_i|^2/4}
\end{gathered}
\]
Substituting this state into the correlation function we obtain  
\[
C^{(2)}_{qh} (\eta,\eta')=\frac{\tr\rho_{\eta'}\rho_\eta}{\tr\rho^2}\sim \frac{1}{|\eta-\eta'|^{\alpha^2/m}} \,,
\label{eq:Renyi2_3}
\] 
where $\alpha$ is the exponent appearing in \eqnref{eq:Renyi2_2}.
In contrast to previously studied examples of decoherence induced transitions in topological states~\cite{bao2023mixed}, here the anyons do not fully condense even at infinite decoherence. The reason is that they are charged under the global $U(1)$ symmetry, and the density dephasing preserves the original probability distribution in position space.

\subsection{Critical filling for the von Neumann quantities\label{sec:laughlin_vN}}

It is straightforward to generalize the analysis of the previous section to R\'enyi-$n$ quantities with $n=2,3,4,\cdots$. In this case the decoherence-induced transition is described by the statistical mechanics of $n$ layers of coupled plasmas. All the relative plasma modes exhibit a BKT transition below a critical filling $\nu_c$ that depends on the R\'enyi index: 
\begin{equation}
    \nu_c(n) = \frac{n-1}{2n},~n=2,3,\cdots\,.
    \label{eq:critical filling for n}
\end{equation}
Quantities that have direct information-theoretic meanings, and bear directly on the recoverability of quantum information, require taking the replica limit $n\rightarrow 1$ ~\cite{Schumacher:1996dy,lessaSWSSB,liu2024wightman,weinstein2025efficient}. 
However, we show in appendix \ref{app:replica_limit} that $\nu_c(n)$ is not analytic at $n = 2$ because taking the limit $n\to 1$ does not commute with taking the thermodynamic limit.
Consequently, the expression \eqref{eq:critical filling for n}, derived for integer $n\ge 2$, does not extend to $n<2$. 
In \appref{app:replica_limit}, we derive the correct expression for non-integer $n$ and prescribe the proper way for taking the $n\to 1$ limit. 

Here, we present a simpler approach, working directly with the R\'enyi-1 or von Neumann quantities. As in the R\'enyi-2 case, we take the fully decohered state $\rho_{\infty}$ as a starting point and show that it is critical also at $n=1$. Then we assess the stability of the critical state to a small deviation from the infinite decoherence limit. This allows us to bound the critical filling factor within the range $1/8 \leq \nu_c(1) \leq 1/4$.

To establish the critical nature of the fully decohered state $\rho_{\infty}$ in the $n \rightarrow 1$ limit, it is convenient to calculate the R\'enyi-1 correlation function for electrons:
\begin{equation}
    C_e^{(1)}(\eta,\eta')=\tr  \big( c(\eta')\sqrt{\rho}c^\dagger (\eta')c(\eta)\sqrt{\rho}c^\dagger(\eta) \big)\,.
\label{eq:Renyi1_0}
\end{equation}
In the electron occupation basis $\ket{\{z\}} = \prod_{i=1}^N c^\dagger(z_i)|0\rangle$, $\rho_\infty$ is fully diagonal with the matrix element: 
\begin{equation*}
    \langle \{z_i\}| \rho_{\infty} |\{z'_i\} \rangle = \frac{1}{Z} |\Psi(\{z_i\})|^2\, \delta_{\{z_i\},\{z'_i\}}\,,
\end{equation*}
where $Z$ is a normalization factor. 
It is then straightforward to calculate $\sqrt{\rho_\infty}$ and the R\'enyi-1 correlation function, for which we have
\begin{equation}
\begin{aligned}
    C_e^{(1)} & (\eta,\eta') \\
    =& \frac{Z^{-1}}{(N-1)!}\int\prod_i^{N-1} d^2z_i |\Psi(\{z_i,\eta\})\Psi(\{z_i,\eta'\})|\,,
\end{aligned}
\label{eq:Renyi1_1}
\end{equation}
where we introduce the notation 
\begin{equation*}
    \{z_i,\eta\} := (z_1,z_2,\ldots,z_{N-1},z_{N}=\eta)
\end{equation*}
to denote a set of coordinates with the last one being specified.
Using the same steps as the analysis for \eqnref{eq:C2e}, we obtain 
\[
C_e^{(1)}(\eta,\eta')\sim \frac{1}{|\eta-\eta'|^{m/2}}, \quad (\mu=+\infty)\,.
\label{eq:Renyi1_2}
\]
We will frequently quote this result in our subsequent analysis of the instability.

As in the R\'enyi-2 case, we assess the stability of the critical phase by expanding the R\'enyi-1 or von Neumann quantities in the small deviation from full decoherence. The convergence of this expansion determines the stability of the critical phase. The same kind of analysis is valid more generally and we use it also to understand the Moore-Read state under decoherence in section \secref{sec:MR_vN}. Hence we will keep the following discussion general and specialize to the Laughlin states toward the end of it.

Specifically, we use $a,b,c$ to label the electron occupation basis states and write the density matrix near the infinite decoherence limit up to the first non-trivial order of the expansion as
\begin{equation}
\begin{gathered}
\rho = \rho_\infty + e^{-\bar{\mu}} \delta \rho + \calO(e^{-2\bar{\mu}})\,, \\
\rho_\infty = \frac{1}{Z} \sum_a |\psi_a|^2 \ket{a} \bra{a}\,,\, 
\delta\rho = \frac{1}{Z} \sum_{(b, c)} \psi_b \psi_c^* \ket{b} \bra{c}\,,
\end{gathered}
\label{eq:large_mu_expansion}
\end{equation}
where $\psi_{a,b,c}$ are the many-body wave functions, $e^{-\bar{\mu}}$ the control parameter of the perturbative expansion.
Here $(b,c)$ enumerates pairs of electron configurations that only differ in one particle:
$$
\bigl\{ (b,c) \,|\, b =\{z_i,\eta\},~c=\{z_i,\eta'\}\bigl\}\,,
$$
so that $\delta\rho$ collects the leading contribution in the off-diagonal coherences. 

Without loss of generality, we choose the von Neumann entropy $S(\rho) = - \tr \rho \log \rho$ and consider its expansion in $\delta \rho$ only up to the first non-trivial order. Analyzing other von Neumann quantities yields the same conclusion.
To properly expand $S(\rho)$ with respect to $e^{-\bar\mu}$, we use the following identity of a positive matrix
\begin{equation}
	\log \rho = \int_0^{+\infty} \big( \frac{1}{s+1} - \frac{1}{s+\rho} \big) ds\,,
\end{equation}
which leads to
\begin{equation}
\begin{gathered}
S(\rho) \approx S(\rho_\infty) + e^{-2\bar\mu}\delta S\,,\\
\delta S = -\int_0^{+\infty} \tr [\delta\rho(s+\rho_\infty)^{-1}\delta\rho(s+\rho_\infty)^{-1}]ds\,.
\end{gathered}
\end{equation}
The fact that $\rho_\infty$ is diagonal in the chosen basis simplifies the calculation, leading to
\begin{equation}
	\delta S = - \frac{2}{Z} \sum_{(b,c)} \frac{|\psi_b| |\psi_c|}{|\psi_c|/|\psi_b| - |\psi_b|/|\psi_c|} \log \frac{|\psi_c|}{|\psi_b|}\,.
\label{eq:deltaS_vN}
\end{equation}
Using the inequality $x/\sinh x \leq 1$ we can find an upper bound on $|\delta S|$ 
\begin{equation}
	\text{(upper bound): }
    |\delta S| \leq \frac{1}{Z} \sum_{(b,c)} |\psi_b| |\psi_c|\,.
    \label{eq:upper bound}
\end{equation}
To derive a lower bound, we first use the inequality $x/\sinh x \geq 1/\cosh x$, which gives
\begin{equation*}
    |\delta S| \geq \frac{2}{Z} \sum_{(b,c)} \frac{|\psi_b| |\psi_c|}{|\psi_c|/|\psi_b| + |\psi_b|/|\psi_c|} \,.
\end{equation*}
We then apply Jensen's inequality $\overline{x^{-1}} \geq \overline{x}^{-1}$ by interpreting $|\psi_b| |\psi_c|$ as being proportional to the weight of the distribution and $|\psi_b|/|\psi_c| + |\psi_c|/|\psi_b|$ in the denominator as the observable. This leads to
\begin{equation*}
    |\delta S| \geq \frac{1}{Z}
    \frac{\big( \sum_{(b,c)}|\psi_b| |\psi_c| \big)^2}{\sum_{(b,c)}|\psi_b|^2 + |\psi_c|^2 }\,.
\end{equation*}
Finally, using $\sum_{(b,c)}|\psi_b|^2 + |\psi_c|^2 \approx 2NL^2Z$ yields
\begin{equation}
\text{(lower bound): }
    |\delta S| \geq \frac{1}{2N L^2 Z^2}
    \big( \sum_{(b,c)}|\psi_b| |\psi_c| \big)^2\,.
    \label{eq:lower bound}
\end{equation}
Using the formulae \eqref{eq:upper bound} and \eqref{eq:lower bound}, we have
\begin{equation}
\begin{aligned}
    \frac{1}{2NL^2} \big(\int d^2 \eta d^2 \eta' & C_e^{(1)}(\eta,\eta') \big)^2 \\
    & \leq |\delta S| \leq \int d^2 \eta d^2 \eta' C_e^{(1)}(\eta,\eta')\,,
\end{aligned}
\label{eq:bound_DeltaS}
\end{equation}
where $C_e^{(1)}$ is exactly the R\'enyi-1 correlation function we have encountered in \eqnref{eq:Renyi1_1}. 

The physical instability comes from unbound pairs separated over a macroscopic distance of order the system size $|\eta-\eta'|\sim O(L)$. 
We can bound the contribution of such macroscopically separated pairs, which we denote as $\delta S_L$, by using the result \eqnref{eq:Renyi1_2} for the R\'enyi-1 correlations in the Laughlin state. On one side of the inequality, we have
\begin{equation}
    |\delta S_L| \leq C L^{4-m/2}
\end{equation}
where $C$ is some constant. Thus, for $m>8$ the correction is convergent in the IR and there is no unbinding instability. In other words the critical filling $\nu_c\geq 1/8$. 
The other side of the inequality gives
\begin{equation}
    |\delta S_L| \geq C' L^{4-m}
\end{equation}
where  $C'$ is some other constant. This inequality implies that there must be an instability for $m \leq 4$, therefore the critical filling $\nu_c \leq 1/4$.

\section{Diagnosing quantum memory\label{sec:laughlin_memory}}

In this section, we determine how the decoherence induced phase transition in the Laughlin state, identified using the plasma analogy, impacts the quantum memory encoded in the degenerate ground states on a torus. 
These degenerate ground states differ only in non-local properties and are therefore complicated to represent  using the plasma analogy. 
Instead we turn to an effective field theory description, which also sets the stage for analyzing the Moore–Read state in the next section.

\subsection{Quantum coherent information}

The quantum coherent information quantifies the quantum information that is recoverable by an optimal decoder after the action of decoherence.
Below we briefly review the definition of this quantity and discuss its general form in the context of Laughlin states.

Let $|\Psi_\ell\rangle, \ell = 0,\ldots,m-1$, denote the $m$-fold degenerate \emph{normalized} Laughlin states on the torus.
We initialize the system $Q$ into a maximally entangled state with a reference qudit $R$
\begin{equation*}
    |\Psi_{RQ}\rangle = \frac{1}{\sqrt{m}}\sum_{\ell=0}^{m-1} |\Psi_{\ell}\rangle\otimes |\ell\rangle_R\,. 
\end{equation*}
Let $\rho_{RQ}$ denote the state after decoherence, the quantum coherent information is defined as
\begin{equation}
    I_c := S(\rho_{Q})-S(\rho_{RQ}),
\end{equation} 
where $S(\rho)$ is the von Neumann entropy.
The quantum coherent information is maximal $I_c = \log m$ for the initial state $\ket{\Psi_{RQ}}$, and it cannot increase under arbitrary quantum channels.
Its value is bounded by $-\log m\leq I_c\leq \log m$. 
We can perfectly recover the encoded information if and only if $I_c=\log m$.
Here, for simplicity, we only consider the R\'enyi-2 version of the quantum coherent information
\begin{equation}
    I_c^{(2)} = \log \frac{\tr  \rho_{RQ}^2}{\tr \rho_Q^2}\,.
\end{equation}
Besides its computational simplicity, this R\'enyi-2 version also has an information-theoretical meaning: $I_c^{(2)} > 0$ implies that $Q$ and $R$ are non-separable, a necessary condition for the decodability.\footnote{Any separable state across $R$ and $Q$ can be written as a convex sum $\rho_{QR} = \sum_k p_k |Q_k\rangle\langle Q_k| \otimes |R_k\rangle\langle R_k|\,, p_k >0$, where $|Q_k\rangle$ and $|R_k\rangle$ are sets of generically nonorthogonal but normalized pure states on $Q$ and $R$. It is then easy to see that $\tr \rho_{QR}^2 \leq \tr \rho_Q^2$ and $I_c^{(2)}$ cannot be positive.}
From now on, we focus on the R\'enyi-2 coherent information and determine how it evolves across the decoherence induced transition identified in the previous section.

It is convenient to use the doubled space description, where the density matrix is vectorized,
\[
\rho=\sum_{a,b}\rho_{ab}|a\rangle\langle b|\rightarrow |\rho\rrangle=\sum_{ab}\rho_{ab}|a\rangle\otimes|b^*\rangle\,.
\]
Here $|b^*\rangle$ denotes the complex conjugate wave function in the density basis.
We can view the pure state $|\rho\rrangle$ as a bilayer system, where the two layers correspond to the ket and bra space. 
According to our definition of the complex conjugation, the density matrix of the undecohered Laughlin state is mapped to a bilayer of Laughlin states with opposite chiralities.
In the doubled space, the decoherence channel is mapped to an operator
\begin{equation}
    \calN = e^{\mathcal{L}t} = e^{-\frac{\gamma t}{2}\int d^2 z (\hat n_1(z)-\hat n_2(z))^2}\,,
\end{equation}
and we have
\begin{equation}
    I_c^{(2)}
    = \log \frac{\sum_{\ell, \ell'=0}^{m-1} \llangle \Psi_{\ell}\Psi^*_{\ell'}|\mathcal{N}^\dagger \mathcal{N}|\Psi_\ell\Psi^*_{\ell'}\rrangle}{\sum_{\ell,\ell'=0}^{m-1} \llangle \Psi_{\ell'}\Psi^*_{\ell'}|\mathcal{N}^\dagger \mathcal{N}|\Psi_\ell\Psi^*_{\ell}\rrangle}\,,
    \label{eq:Ic2 original}
\end{equation}
where the denominator and numerator have subtle but important differences in the indices.
The value of the (R\'enyi-2) quantum coherent information is independent of the choice of the basis states. Below, we simplify the expression by choosing a particular basis.

We can attribute the $m$-fold degeneracy of the Laughlin state to its center-of-mass (CM) motion~\cite{Haldane1985many}.
Thus, we can pick our basis states $\ket{\Psi_\ell}$ to have a definite CM position along the $y$-direction, or equivalently, a definite CM momentum along the $x$-direction.
More explicitly, let $t_i(\mathbf{a})$ denote the magnetic translation operator that moves the $i$-th electron by $\mathbf{a}$, and $T(\mathbf{a})=\prod_it_i(\mathbf{a})$ the CM translation.
We can specify the basis states by 
\begin{equation}
    T_x \ket{\Psi_\ell} = e^{\frac{2\pi i}{m} \ell} \ket{\Psi_\ell}\,,\quad
    \ket{\Psi_{\ell+1}} = T_y \ket{\Psi_\ell}\,,
    \label{eq:magnetic_eig}
\end{equation}
where $T_i=T(\mathbf{e_i} L_i / N_\phi)$, $i=x,y$, moves the CM position along the $i$-th direction by $L_i/m$, and they serve as the logical operators for the code subspace. 
Note that the effect of the dephasing channel does not change if we apply the magnetic translation $T_i$ to the ket and bra state simultaneously, i.e., $[\mathcal{N},T_i\otimes T^*_i]=0$.
We then have the following simplified form:
\begin{equation}
    I_c^{(2)} = \log \frac{\sum_{\ell=0}^{m-1}{\llangle \Psi_\ell\Psi^*_{0}|\mathcal{N}^\dagger \mathcal{N}|\Psi_\ell\Psi^*_{0}\rrangle}}{\sum_{\ell=0}^{m-1}\llangle \Psi_{0}\Psi^*_{0}|\mathcal{N}^\dagger \mathcal{N}|\Psi_\ell\Psi^*_{\ell}\rrangle}\,,
\label{eq:Ic_2}
\end{equation}
where we reduce the double summations in \eqnref{eq:Ic2 original} to single ones.

\subsection{Effective field theory of the decohered Laughlin states\label{sec:Laughlin_eff}}

Under the plasma analogy, degenerate Laughlin states correspond to configurations with magnetic and electric fluxes inside the torus, which complicates direct statistical analysis. 
In this section, we reformulate the bilayer plasma discussed in the previous section in terms of an effective field theory, leveraging the established correspondence between Laughlin wavefunctions and (1+1)-dimensional conformal field theories (CFTs)~\cite{moore1991Nonabelions}. 
This equivalent framework provides a more transparent description of the degenerate ground states, facilitating the analysis of quantum coherent information.

To represent the $\nu = 1/m$ Laughlin wave function, we consider the $U(1)_m$ compact boson CFT
\begin{equation}
    S_0=\frac{m}{8\pi}\int d^2z (\nabla\phi)^2\,,\quad \phi \simeq \phi + 2\pi\,.
    \label{eq:S0 1/m Laughlin state}
\end{equation}
The bosonic field $\phi$ can be decomposed into chiral and anti-chiral components as $\phi(z,\bar{z}) = \varphi(z)+\bar\varphi(\bar z)$.
The Laughlin wave function on the torus can be expressed in terms of the correlation function of the chiral vertex operators~\cite{moore1991Nonabelions}. Specifically, on an $L_x \times L_y$ torus, we write down an \emph{un-normalized} Laughlin wave function as 
\begin{equation}
    \Psi_\ell(\{z_i\}) = \tr_{\mathcal{H}_\ell} \big( e^{-L_y H_\varphi} \prod_{i=1}^N e^{im\varphi(z_i)}e^{-\frac{i}{2\pi}\int d^{2} z\varphi} \big)\,,
\label{eq:torus_laughlin}
\end{equation}
where $H_\varphi$ is the Hamiltonian of the chiral boson, and $\mathcal{H}_l$ denotes the Hilbert space with twisted boundary conditions of the chiral field along the $x$-direction
\begin{equation}
    \varphi(z + L_x) = \varphi(z) + 2\pi \big( k + \frac{\ell}{m} \big)\,,\quad k \in \bbZ\,.
    \label{eq:compact boson pbc}
\end{equation}
Importantly, we do not normalize the right-hand side of \eqnref{eq:torus_laughlin} by the $\ell$-dependent chiral character $\chi_\ell= \tr_{\mathcal{H}_\ell}e^{-L_yH_\varphi}$ so that the resulting wave functions satisfy \eqnref{eq:magnetic_eig}. Their overlaps take the form
\begin{equation}
\langle \Psi_\ell|\Psi_{\ell'}\rangle=C\delta_{ll'}\,,
\label{eq:psi ell inner product}
\end{equation}
where $C$ is an $\ell$-independent positive constant.
We therefore adopt the convention \eqref{eq:torus_laughlin} in the subsequent analysis.
In the following, we use $\langle \cdots\rangle'$ to denote such un-normalized correlation functions and use $\langle\cdots\rangle$ for the normalized expectation values.

Using this representation, we can express the wave function overlaps in the numerator and denominator of \eqnref{eq:Ic_2} as path integrals of two non-chiral fields. 
More generally, we consider
\begin{equation}
\begin{aligned}
    Z(\bar{\ell}_1,\ell_2; \ell_1, \bar{\ell}_2) = \llangle \Psi_{\bar{\ell}_{1}} \Psi^*_{\ell_{2}}| \mathcal{N}^\dagger \mathcal{N}|\Psi_{\ell_{1}} \Psi^*_{\bar{\ell}_{2}}\rrangle\,.
\end{aligned}
\end{equation}
We need two chiral fields $\varphi_1$, $\varphi_2$ for $\ket{\Psi_{\ell_1}}$, $\bra{\Psi_{\ell_2}^*}$, and two anti-chiral fields $\bar{\varphi}_1$, $\bar{\varphi}_2$ for $\bra{\Psi_{\bar{\ell}_1}}$, $\ket{\Psi_{\bar{\ell}_2}^*}$, which should obey the following spatial boundary conditions
\begin{equation}
    \centering
    \renewcommand{\arraystretch}{1.6}
    \setlength{\tabcolsep}{4.pt}
    \begin{tabular}{c|c|c|c|c}
    \hline
    & $\bar{\varphi}_1$& $\varphi_2$& $\varphi_1$& $\bar{\varphi}_2$\\
          \hline
          winding & $\bar{k}_1 - \frac{\bar{\ell}_1}{m}$&  $k_2 + \frac{\ell_2}{m} $& $k_1+\frac{\ell_1}{m}$& $\bar k_2 - \frac{\bar{\ell}_2}{m}$ \\
 \hline 
 \end{tabular}
 \label{tab:winding}
\end{equation}
We can combine them into non-chiral fields $\phi_\alpha = \varphi_\alpha + \bar{\varphi}_\alpha$ and write the overlap as 
\begin{equation}
\begin{aligned}
    Z(\bar{\ell}_1,\ell_2; \ell_1, \bar{\ell}_2) = & \int Dn_{1,2} D\phi_{1,2}\, e^{-\gamma t\int (n_1-n_2)^2} \\
    & \times e^{-\sum_{\alpha=1}^2S_0[\phi_\alpha]+im\int (n_\alpha-\bar n)\phi_\alpha}\,,
\end{aligned}    
\label{eq:Z l expression}
\end{equation}
where $n_\alpha$ is the electron density and $\bar{n} = 1/2\pi m$ ensures the correct total electron number.

As a sanity check, we examine \eqnref{eq:Z l expression} at $\mu = 0$, where $n_1,\phi_1$ and $n_2,\phi_2$ correspond to the two decoupled single-component plasmas as described in \eqnref{eq:purity_Laughlin}. 
The fact that each plasma is screening (for $m < 70$) implies that both $\phi_1$ and $\phi_2$ are gapped fields.
The screening behavior further indicates that magnetic charges are linearly confined. Consequently, threading a magnetic flux through the torus incurs an extensive free energy cost, which leads to the following no-winding condition for the two non-chiral fields:
\begin{equation}
    k_\alpha + \bar{k}_\alpha + \frac{\ell_\alpha - \bar{\ell}_\alpha}{m} = 0\,,\quad \alpha = 1,2\,.
\end{equation}
Thus we recover \eqnref{eq:psi ell inner product}.

This simple case illustrates the general strategy that we adopt in the subsequent analysis, including the treatment of the Moore–Read state later.
Specifically, we start with kinematic analysis with a field theory, e.g., \eqref{eq:Z l expression}, then invoke the screening property from the plasma analogy to determine the low-energy degrees of freedom. 
With this input, we carry out the remainder of the analysis with the field-theory technique. For example, the confinement of magnetic charge is not immediately evident from the screening property due to its non-local formulation in terms of electric charges. 
The corresponding no-winding condition emerges naturally as a consequence of the fields being gapped.

We close this section with a remark on the generality of our results. Our analysis so far is fully based on the model wave function of the Laughlin states. The effective field theory approach may have broader applicability.
The model wave function can also be obtained at the temporal boundary of the Euclidean-time path integral of $U(1)_m$ Chern-Simons field theory, where the states $|\Psi_\ell\rangle$ are distinguished by the holonomy of the gauge field $a$ around the non-contractible $x$-cycle 
\begin{equation}
    \oint_{\gamma_x}a = 2\pi \big( k + \frac{\ell}{m} \big)\,,\quad k \in \bbZ\,.
\end{equation}
Deviation from the model wave function can be interpreted as arising from the inclusion of gapped matter fields coupled to the Chern-Simons gauge field. By appealing to the principle of universality, we expect Eq.\eqref{eq:Z l expression} to retain the dominant contributions relevant for long-wavelength behavior. Accordingly, we regard Eq.~\eqref{eq:Z l expression} as an effective field theory for more general Laughlin states subject to density dephasing.

\subsection{Filling-dependent decodability transition}

In the plasma analogy, it is convenient to reorganize the bilayer plasma into a paired plasma and a ``relative plasma".
Accordingly, we re-combine the field variables as follows
\begin{equation}
    n_\pm = n_1 \pm n_2\,,\, \phi_\pm = \frac{1}{2} (\phi_1 \pm \phi_2)\,,\,\theta_\pm = \theta_1 \pm \theta_2\,,
\end{equation}
where $\theta_\alpha = m(\varphi_\alpha - \bar{\varphi}_\alpha)/2$, $\alpha = 1,2$ are the $2\pi$-periodic non-chiral field dual to $\phi_\alpha$. It is necessary to introduce them to keep track of the boundary conditions faithfully.
We then recast the partition function as
\begin{equation*}
\begin{gathered}
    Z(\bar{\ell}_1,\ell_2;\ell_1,\bar{\ell}_2) = \!\! \int \!\calD (n_\pm,\phi_\pm) e^{-S_+[\phi_+, n_+] - S_-[\phi_-,n_-]}\,, \\
    S_+ [\phi_+, n_+] = 2 S_0[\phi_+] - im \!\int \big(n_+ - 2\bar{n}) \phi_+\,, \\
    S_-[\phi_-, n_-] = 2 S_0[\phi_-] - \!\int im n_- \phi_- + \gamma t n_-^2\,.
\end{gathered}
\end{equation*}
Namely, $n_+, \phi_+$ correspond to the charge-$2\sqrt{m}$ paired plasma with a uniform background charge, and $n_-,\phi_-$ correspond to the ``relative" plasma with a decoherence-controlled fugacity. 

Despite the formal separation into $S_+$ and $S_-$, the fields $n_+,\phi_+$ and $n_-,\phi_-$ are not entirely independent in general.
Recall that the paired plasma is in the screening phase (for $m<35$), thus implying that $\phi_+$ remains gapped regardless of the decoherence strength.
Consequently, energetically favored configurations of $\phi_+$ should obey the no-winding condition, which in turn constrains the spatial boundary conditions for $\phi_-$ and $\theta_-$:
\begin{equation}
\begin{gathered}
    \frac{k_1+\bar k_1+k_2+\bar k_2}{2}+\frac{\ell_1-\bar\ell_1+\ell_2-\bar \ell_2}{2m}=0\,, \\
    \begin{aligned}
        \phi_-(z + L_x) =& \phi_-(z) + \big(k_1 + \bar{k}_1 + \frac{\ell_1 - \bar{\ell}_1}{m} \big) 2\pi\,, \\
        \theta_-(z + L_x) =& \theta_-(z) + m\big(k_1 + \bar{k}_2 + \frac{\ell_1 - \bar{\ell}_2}{m} \big) 2\pi \,.
    \end{aligned}
\end{gathered}
\label{eq:w_+=0}
\end{equation}
Furthermore, to account for the quantized charge of the underlying particles, we need to introduce a short-distance cutoff and discretize the space into a lattice so that the density fields $n_+$ and $n_-$ become integer-valued fields defined on lattice sites. Under this discretization, it becomes clear that $n_+$ and $n_-$ are not independent either, as their sum must always yield even integers.

Since $\phi_+$ remains gapped, the presumed decodability transition should be governed by the dynamics of $\phi_-$.
To isolate this, we integrate out $n_+, \phi_+$ and write
\begin{equation}
\begin{gathered}
    Z(\bar{\ell}_1,\ell_2; \ell_1, \bar{\ell}_2) = Z_\infty Z_-(\ell_1 - \bar{\ell}_1, \ell_1 - \bar{\ell}_2)\,, \\
    Z_\infty := \int \calD(n_+,\phi_+) e^{- S_+[n_+, \phi_+]} \Big{|}_{n_- = 0}\,, \\
    Z_-(\ell_1 - \bar{\ell}_1, \ell_1 - \bar{\ell}_2) := \int \calD(n_-,\phi_-) e^{- S_{\text{eff}}[n_-, \phi_-]}\,,
\end{gathered}
\label{eq:Z separation}
\end{equation}
where $Z_\infty$ is obtained by integrating out $n_+,\phi_+$ at fixed $n_- = 0$; $Z_-$ encodes the dynamics of $\phi_-$, which contains a summation over allowed winding numbers of $\phi_-,\theta_-$ subjected to the no-winding condition \eqref{eq:w_+=0}.
We can compare \eqnref{eq:Z separation} with \eqnref{eq:laughlin_near_infty1} and recognize that $Z_\infty$ corresponds to the paired plasma at the strong decoherence limit, and $Z_-$ to the ``relative" plasma.  \eqnref{eq:laughlin_near_infty1} in the plasma analogy holds only near the strong decoherence limit, while \eqnref{eq:Z separation} and the following analysis can be applied more broadly.

To determine the structure of $S_{\text{eff}}[n_-,\phi_-]$, we note that integrating out $n_+,\phi_+$ generates terms involving $n_-$ alone, as $n_-$ is the only field locally coupled with them.
Moreover, the original action is symmetric under $n_- \rightarrow -n_-$ and $\phi_- \rightarrow -\phi_-$, implying that only even powers of $n_-$ can appear. 
Performing an expansion in the powers of $n_-$ and its gradients, we have
\begin{equation}
    S_{\text{eff}}[n_-, \phi_-] = 2 S_0[\phi_-] -\! \int imn_-\phi_- + \bar\gamma t n_-^2 + \ldots   
    \label{eq:Seff n- phi-}
\end{equation}
The coefficient $\bar\gamma t$ implicitly includes the renormalization from integrating out $n_+, \phi_+$, and the ellipsis denotes terms involving gradients of $n_-$ or higher powers.
Near the strong decoherence limit, $n_-$ only weakly fluctuates around $n_- = 0$ so that it suffices to omit the subleading terms in \eqnref{eq:Seff n- phi-} and further integrate out $n_-$. 
A proper treatment requires us to discretize the space into a lattice with a short-distance cutoff $a$. 
This procedure yields a sine-Gordon field theory 
\begin{equation}
\begin{gathered}
    Z_-(\ell_1 - \bar{\ell}_1, \ell_1 - \bar{\ell}_2) = \int \calD \phi_- e^{-S_{\text{eff}}[\phi_-]}\,, \\
    S_{\text{eff}}[\phi_-] \approx \int \frac{m}{4\pi}(\nabla\phi_-)^2 - 2e^{-\bar\mu}\cos m\phi_-\,,
\end{gathered}
\label{eq:SG}
\end{equation}
Since the boundary conditions involve the dual field $\theta_-$, it is convenient to recast \eqnref{eq:SG} with the corresponding Hamiltonian that includes $\theta_-$ explicitly
\begin{equation}
\begin{gathered}
    Z_-(\ell_1 - \bar{\ell}_1, \ell_1 - \bar{\ell}_2) \approx \tr_{\ell_1 - \bar{\ell}_1, \ell_1 - \bar{\ell}_2} e^{-L_y H_-}\,, \\
    H_-=\!\!\int_0^{L_x} \frac{(\partial_x\theta_-)^2}{4\pi m} + \frac{m}{4\pi} (\partial_x\phi_-)^2 - 2e^{-\bar\mu}\cos m\phi_-\,.
\end{gathered}
\label{eq:H_SG}
\end{equation}
Even though we need approximations that are well-controlled only near the strong decoherence, we expect \eqnref{eq:SG} and \eqref{eq:H_SG} to capture the possible phases and the potential transition away from this limit.

Using \eqnref{eq:Z separation}, we can write the R\'enyi-2 quantum coherent information in terms of $\phi_-$ only:
\begin{equation}
    I_c^{(2)}=\log \frac{\sum_{\ell=0}^{m-1} Z_-(0,\ell)}{\sum_{\ell=0}^{m-1} Z_-(\ell,0)}\,.
\label{eq:Ic2_Z}
\end{equation}
It is also convenient to divide both the numerator and denominator by $Z_-(0,0)$ and write
\begin{equation}
I_c^{(2)}=\log \frac{1+\sum_{\ell=1}^{m-1} e^{-\Delta F_{(0,\ell)}}}{1+\sum_{\ell=1}^{m-1}e^{-\Delta F_{(\ell,0)}}}\,,
\label{eq:Ic2 Delta F}
\end{equation}
where $\Delta F_{(0,\ell)}$ and $\Delta F_{(\ell,0)}$ are the free energy costs of twisting the boundary conditions of $\theta_-$ and $\phi_-$.
At the maximal dephasing $\mu=\infty$, the sine-Gordon theory is reduced to the compact boson CFT. The duality symmetry $\phi_-\leftrightarrow \theta_-/m$ guarantees that $Z_-(0,\ell) = Z_-(\ell,0)$, and thus $I_c^{(2)}=0$.  This result is expected because the fully decohered state becomes separable between the system $Q$ and reference $R$, which cannot contain any positive coherent information.
At finite $\mu$, the duality symmetry is explicitly broken, leading to nonzero $I_c^{(2)}$.
There are two phases, in which $I_c^{(2)}$ exhibits distinct behavior:
\begin{enumerate}
\item The topologically ordered encoding phase is established for $\mu<\mu_c$. In this case the cosine interaction is strong enough to gap out $\phi_-$ and energetically penalize its spatial winding.
Accordingly, we have $\Delta F_{(0,\ell)} \approx 0$ while $\Delta F_{(\ell \neq 0,0)} \sim  e^{-\bar{\mu}} L$. Therefore $I_c^{(2)}=\log m$ in the thermodynamic limit. 

For $m\leq 4$, the critical decoherence strength $\mu_c$ is pushed to infinity in the thermodynamic limit. In this case $I_c^{(2)}=\log m$ for any finite decoherence strength or time. 
In a finite-size system, on the other hand, $I_c^{(2)}$ will start to deviate from $\log m$ if the free energy cost $\Delta F_{(\ell \neq 0,0)}$ becomes negligible. 
This happens when $e^{\bar{\mu}} L \sim \calO(1)$, or equivalently, when $\bar\mu \sim \calO(\log L)$. 
Since $\bar\mu$ is linear in the decoherence time $t$, the lifetime of the quantum memory is therefore of order $\calO(\log L)$.
It is worth pointing out that the scaling of the lifetime with $L$, in this case, saturates the maximum that is achievable under such density dephasing~\cite{sang2025stability}.

\item The critical phase is established for $\mu_c\leq \mu<\infty$.  In this case the cosine term is irrelevant, so we can analyze its contribution perturbatively. Surprisingly, it contributes an additional $O(1)$ free energy cost for nonzero windings of $\phi_-$, i.e., $\Delta  F_{(\ell,0)} >\Delta F_{(0,\ell)} (\ell\neq 0)$, which leads to $0<I_c^{(2)}<\log m$. The calculation is shown in \appref{app:perturbation}.
\end{enumerate}
In Fig.~\ref{fig:intro fig 2}~(a), we plot the R\'enyi-2 coherent information as a function of decoherence strength of the Laughlin states (for both $m>4$ and $m\leq 4$) under dephasing. Remarkably, for $m\leq 4$, the quantum memory is robust for any finite dephasing noise in contrast to that of the toric code under $Z$ dephasing. For $m>4$, the coherent information remains finite and decrease continuously after the transition, meaning that the critical phase has partial quantum memory. This is in sharp contrast to the toric code under dephasing noise, where the quantum memory is completely destroyed after the transition.

\section{Decohered Moore-Read States}
\label{sec:Moore Read}

In this section, we analyze the decohered Moore-Read (MR) states, the simplest non-Abelian quantum Hall states~\cite{moore1991Nonabelions}. 
The most prominent feature of the MR states is the non-Abelian quasi-hole excitations. 
Multiple well-separated quasi-holes can span a multi-dimensional fusion space that can be used for storing quantum information and performing topological quantum computation.
The main goal of this section is to examine the integrity of this fusion space under decoherence. We again primarily consider R\'enyi-2 quantities and discuss the von Neumann quantities at the end.

\subsection{Review of the Moore-Read wave function}
\label{sec:MR}

The MR wave function at $\nu =1/m$ is constructed from the chiral Ising and chiral boson CFT.
Let $\psi$ be the chiral fermion field in the Ising sector and $\varphi$ the chiral boson field with $\langle \varphi(z)\varphi(w)\rangle=-\frac{1}{m}\log (z-w)$. 
The electron corresponds to $\psi e^{im\varphi}$, and the (un-normalized) MR wave function of $N$ electrons on the plane reads
\begin{equation}
    \Psi(\{z_i\})=\langle \prod_{i=1}^N\psi(z_i)e^{im\varphi(z_i)} e^{-\frac{i}{2\pi }\int d^2 z \varphi}\rangle, 
    \label{eq:MR wave function def}
\end{equation}
where we need to assume $N$ is even and $e^{-\frac{i}{2\pi}\int d^2 z \varphi}$ provides the neutralizing background charge.
For simplicity, we introduce $\langle \cdots\rangle_{\text{bg}}$ as the expectation value with the background charge.
If there are non-chiral or multi-component fields $\{\phi_\alpha\}$, we use the same notation,
\begin{equation}
\langle \cdots\rangle_{\text{bg}}=\langle \cdots e^{-\frac{i}{2\pi}\sum_\alpha\int d^2 z\phi_\alpha}\rangle\,.
\end{equation}
The reader can infer the correct background charge from the context.
The MR state hosts $3m$ types of anyons corresponding to the following primary fields
\begin{equation}
\{e^{i\ell\varphi},\psi e^{i\ell\varphi},\sigma e^{i(\ell+\frac{1}{2})\varphi}\}\,, \, \ell=0,1,\cdots,m-1\,,
\end{equation}
all of which are local with respect to the operator $\psi e^{im\varphi}$.
For instance, $e^{i\varphi}$ creates a Laughlin quasi-hole carrying charge $1/m$, while $\sigma e^{i\varphi/2}$ corresponds to the fundamental quasi-hole with charge $1/2m$.
Two fundamental quasi-holes can fuse into more than one particles
\begin{equation}
    \sigma e^{i\varphi/2}\times \sigma e^{i\varphi/2}=e^{i\varphi}+\psi e^{i\varphi},
\end{equation}
and thus they are non-Abelian anyons. In the following discussion, ``quasi-hole" always refers to the fundamental quasi-hole unless specified explicitly.

When there are multiple quasi-holes, their wave functions span a degenerate subspace called the fusion space. 
As the simplest example, having four quasi-holes at locations $\eta_1,\eta_2,\eta_3,\eta_4$ yields two basis states
\begin{equation}
\begin{aligned}
\Psi_{\mu,\{\eta_k\}} & (\{z_i\}) = \\
\mathcal{F}_\mu & (z_i;\eta_k)\times \langle\prod_{k=1}^4e^{i\varphi(\eta_k)/2} \prod_{i=1}^N e^{im\varphi(z_i)}\rangle_{\text{bg}}\,,
\end{aligned}
\label{eq:four_qhs}
\end{equation}
where $\mu = 1, \psi$ labels the fusion outcome of the first two $\sigma$'s, and $\mathcal{F}_1$, $\mathcal{F_\psi}$ are the two chiral conformal blocks of the following correlation function of non-chiral primaries~\cite{Belavin1984}
\begin{equation}
\begin{aligned}
\langle \prod_{k=1}^4 \sigma(\eta_
k,\bar \eta_k)\prod_{i=1}^N \varepsilon(z_i,\bar z_i)\rangle
& = \\ \sum_{\mu=1,\psi}\mathcal{F_\mu} & (z_i;\eta_k) \overline{\mathcal{F}}_\mu( \bar z_i;\bar \eta_k)\,,
\end{aligned}
\label{eq:foursigma}
\end{equation}
where $\varepsilon = \bar\psi \psi$. When the quasi-holes are far apart, the Gram matrix of the two wave functions is~\cite{bonderson2011Plasma}
\begin{equation}
    \langle\Psi_{\mu,\{\eta_k\}}|\Psi_{\nu,\{\eta_k\}}\rangle = C \delta_{\mu,\nu} + \calO(e^{-|\eta_i - \eta_j|/\xi})\,,
    \label{eq:4qh wave function orthonormal}
\end{equation}
where $C$ is a positive constant that is independent of the fusion channel and $\xi$ is a correlation length.
Namely, they are two orthogonal states. We can use the fusion space to encode one qubit represented by entangling this state with an external reference qubit
\begin{equation}
|\Psi_{RQ}\rangle=\frac{1}{\sqrt{2}}(|\Psi_{1,\{\eta_k\}}\rangle\otimes|0_R\rangle+|\Psi_{\psi,\{\eta_k\}}\rangle\otimes|1_R\rangle)\,.
\label{eq:4qhs_encoding}
\end{equation}
As a proxy for the recoverable quantum information we will compute the R\'enyi-2 quantum coherent information from $R$ to $Q$ of the above state subject to decoherence.  

\subsection{Effective field theory near the strong decoherence limit\label{sec:MR_eff}}

The CFT construction of the Moore-Read wavefunctions serves as a convenient starting point for analyzing the effect of decoherence on encoded quantum information. We begin by calculating the purity, which serves as a generating function for various information quantities and can be directly represented using the wavefunctions \eqref{eq:MR wave function def} in close analogy to the analysis of the decohered Laughlin states.
\begin{equation}
\begin{aligned}
    \tr \rho^2
    =&\int \calD(z^1_i,z^2_i) e^{-V(\{z^1_i\},\{z^2_i\})}\\
    &\quad\langle \prod_{\alpha=1}^2\prod_{i=1}^N\varepsilon_\alpha(z^\alpha_i,\bar z_i^\alpha)e^{im\phi_\alpha(z^\alpha_i,\bar z^\alpha_i)}\rangle_{\text{bg}}\,.
\end{aligned}
\label{eq:purityMR}
\end{equation}
This expression of the purity can be viewed as insertions of specific perturbations in a pair of Ising  and a pair of $U(1)$ CFTs. Here $\varepsilon_\alpha$ denotes the energy operator of the Ising CFT $\alpha$ and $\phi_\alpha=\varphi_\alpha+\bar{\varphi}_\alpha$ is the non-chiral boson field of the corresponding $U(1)$ CFT; $V$ represents a local attraction between the two liquids with electron locations given by the complex number  $z^\alpha_i$.
As in the case of Laughlin states, treated above, our strategy for determining the decoherence induced transitions is 
building an effective theory in terms of perturbations from the fully decohered limit.

Before considering the fully decohered state it is worth noting the symmetries intrinsic to the field theory \eqref{eq:purityMR}, which are present at any decoherence strength. 
Many of these symmetries originate from the presence of anyons in the MR state. Here we highlight  two that are particularly useful for constructing the effective theory.
The first symmetry, inherited from the Laughlin quasi-hole, is a $\bbZ_m^{\otimes 2}$ symmetry acting on the $U(1)^{2}$ part alone.
It shifts the boson fields $\phi_{1,2}$ by
\begin{equation}
(\bbZ_m)_i: \,\phi_i \rightarrow \phi_i + \frac{2\pi k_i}{m}\,,\, i=1,2\,,
\label{eq:Zm4 symmetry}
\end{equation}
where $k_i=0,1,\cdots,m-1$. 
The vertex operators $e^{im\phi_{1}}$ and $e^{im\phi_2}$ are clearly invariant.
The extra phase from the background charge can be canceled by adding background vector potential to the corresponding topological defect lines~\cite{Gaiotto2015generalized,chang2019topological}.\footnote{Specifically, the symmetry is generated by the following line operators
\begin{equation}
\begin{aligned}
    U_i(C) = \exp\big( i\frac{k_i}{m}\oint_C (\nabla\theta_i + e \mathbf{A}) \cdot d\ell \big) \,,
\end{aligned}
\label{eq:Uc}
\end{equation}
where $C$ is a loop and $\mathbf{A}$ is the vector potential for the background magnetic field. 
They are related to the worldline of charge-$e/m$ Laughlin quasi-holes.
The partition function $Z_{\text{MR}_\infty}$ with $U_i(C)$ inserted is invariant under continuous deformation of $C$, i.e., the line operators \eqref{eq:Uc} are topological, ensuring the symmetry.
}
The second symmetry, related to the fundamental quasi-holes, is a doubled Kramers-Wannier symmetry that transforms both the $\text{Ising}$ and $U(1)$ fields
\begin{equation}
    \text{KW}_i:\,
    \varepsilon_i\rightarrow -\varepsilon_i\,,\,\phi_i\rightarrow \phi_i+\frac{\pi}{m},\quad  i=1,2 \,.
\end{equation} 
As with the $\bbZ_m^{\otimes 2}$ symmetry, the topological defect line implementing the $\text{KW}^{\otimes 2}$ symmetry must be dressed by the background vector potential as well.

In the strong decoherence limit, the particle positions of the two electron fluids become identified, so that the purity reads
\begin{equation}
    \tr\rho^2_{\infty} = \int \calD \xi \langle \prod_{i=1}^{N}(\varepsilon_1\varepsilon_2)(\xi_i,\bar{\xi}_i)e^{2im\phi_+(\xi_i,\bar{\xi}_i)} \rangle_{\text{bg}}\,,
\label{eq:purity_fullydeMR}
\end{equation}
where $\phi_\pm = (\phi_1 \pm \phi_2)/2$.
It is important to note two sets of symmetries, special to the strong decoherence limit, which will be explicitly broken once we deviate from this limit. One is a $U(1)$ symmetry shifting $\phi_-$ 
\begin{equation}
    U(1): \quad \phi_- \rightarrow \phi_- + \alpha, \quad \alpha \in [0,2\pi)\,,
\end{equation}
which is manifest from the expression \eqref{eq:purity_fullydeMR}.
The other is a $U(1)\times U(1)$ symmetry acting only on the $\text{Ising}^2$ part. To see this, recall that the $\text{Ising}^2$ CFT is equivalent to the $\mathbb{Z}_2$ orbifold of a $2\pi$-periodic compact boson~\cite{Ginsparg:1988ui} 
\begin{equation}
    S_\Phi =\frac{1}{2\pi}\int d^2 z(\nabla\Phi)^2\,,\,\, \Phi\sim \Phi+2\pi,-\Phi\,.
\end{equation}
The latter manifestly has a $U(1)_m\times U(1)_w$ symmetry for $\Phi$ and its dual $\Theta$
\begin{equation}
\begin{aligned}
    U(1)_m: \quad &\Phi\rightarrow \Phi+\alpha_m  ,\quad \alpha_m\in[0,2\pi)\,,\\
    U(1)_w: \quad &\Theta\rightarrow \Theta+\alpha_w ,\quad \alpha_w\in[0,2\pi)\,.
\end{aligned}
\end{equation}
The dictionary of the operator mapping is
\begin{equation}
\begin{aligned}
&\sigma_1\sigma_2 \sim \cos \Phi\,,\quad \mu_1\mu_2\sim\sin \Phi\,,\\
&\sigma_1\mu_2 \sim \cos\frac{\Theta}{2}\,,\quad\mu_1\sigma_2\sim \sin\frac{\Theta}{2}\,, \\
&\varepsilon_{1} + \varepsilon_2 \sim\cos 2\Phi\,,\quad \varepsilon_{1} - \varepsilon_2 \sim\cos \Theta\,,\\
&\varepsilon_1\varepsilon_2\sim (\nabla\Phi)^2 -(\nabla\Theta/2)^2 \,.
\end{aligned}
\label{eq:dictionary}
\end{equation}
One can also read out other global symmetry actions from this dictionary. In particular, the aforementioned $\text{KW}^{\otimes 2}$ symmetries act by 
\begin{equation}
\begin{aligned}
    \text{KW}_1:&\quad 2\Phi \leftrightarrow \pi - \Theta\,, \\
    \text{KW}_2:&\quad 2\Phi \leftrightarrow \Theta\,.
\end{aligned}
\label{eq:KW symmetry boson}
\end{equation}
Since the reference theory \eqref{eq:purity_fullydeMR} only contains the insertion of $\varepsilon_1\varepsilon_2$, the $U(1)_m\times U(1)_w$ symmetry is fully preserved.

Similar to our discussion on the purity of decohered Laughlin state, we organize the deviation from the reference point $Z_{\text{MR}_\infty}\equiv \tr\rho^2_{\infty}$ by the number of unpaired operators
\begin{equation}
\begin{gathered}
    \tr \rho^2 = \sum_{n}e^{-2\bar \mu n} Z_{\text{unpaired}}^{(n)} \,,\quad Z_{\text{unpaired}}^{(0)} = Z_{\text{MR}_\infty}\,.
\end{gathered}
\end{equation}
Near the strong decoherence limit, we can approximate the contribution from unpaired operators as inserting \emph{test} operators to the partition function of the reference theory 
\begin{equation}
\begin{aligned}
    \frac{Z_{\text{unpaired}}^{(n)}}{Z_{\text{MR}_\infty}} \approx & \\
    \frac{1}{(n!)^2}\int \prod_{\alpha=1}^2 & \prod_{i=1}^n d^2 z^\alpha_i
\langle \prod_{\alpha,n} e^{im\phi_\alpha}\varepsilon_\alpha(z^\alpha_i,\bar z^\alpha_i)\rangle_{\text{MR}_\infty}\,,
\end{aligned}
\label{eq:MR Zunpaired}
\end{equation}
where we introduce the following notation for the correlation function in the reference theory
\begin{equation}
\begin{aligned}
\langle & X  \rangle_{\text{MR}_\infty}\equiv \\
&\frac{1}{Z_{\text{MR}_\infty}} \int \calD \xi
\langle X \prod_{p=1}^{N}(\varepsilon_1\varepsilon_2)(\xi_p,\bar\xi_p)e^{2im\phi_+(\xi_p,\bar\xi_p)}\rangle_{\text{bg}}\,.
\end{aligned}
\label{eq:cor_fullydeMR}
\end{equation}
\eqnref{eq:MR Zunpaired} generalizes the approximation \eqref{eq:laughlin_near_infty1} for the decohered Laughlin state in the dilute limit. The inserted test operators $e^{im\phi_\alpha}\varepsilon_\alpha$ preserve the intrinsic symmetries of course, while breaking the symmetries that are special to the fully decohered reference state. 

To construct an effective theory we must identify the low-energy modes of the reference theory and their overlap with the test operators. Equivalently, we need to determine whether any of the symmetries are \emph{spontaneously} broken in the infrared.
For the $\bbZ_m^{\otimes 2}$ symmetry and $U(1)$ shift of $\phi_-$, the plasma analogy yields
\begin{equation}
\begin{aligned}
    \lim_{|\eta|\rightarrow \infty} \langle e^{i\phi_+(\eta,\bar{\eta})} e^{-i\phi_+(0,0)} \rangle_{\text{MR}_\infty} >& 0\,, \\
    \langle e^{i\phi_-(\eta,\bar{\eta})} e^{-i\phi_-(0,0)} \rangle_{\text{MR}_\infty} =& |\eta|^{-1/m}\,.
\end{aligned}
\end{equation}
In the plasma language, inserting $e^{i\phi_+}$ or $e^{i\phi_-}$ amounts to adding two different types of test charge into the plasma. 
The $\phi_+$ charge is completely screened, producing true long-range order, whereas the $\phi_-$ charge is unscreened and shows only quasi-long-range order. 
Consequently, we retain only $\phi_-$ and ignore $\phi_+$ in the effective theory, by making the replacement
\begin{equation}
    e^{im\phi_1} \rightarrow e^{im\phi_-}\,,\quad e^{im\phi_2} \rightarrow e^{-im\phi_-}\,.
\end{equation}
For the $U(1)_m$ and $U(1)_w$ symmetry, note that $\cos \Phi$ translates into test charges in the plasma that are fully screened.
Together with the KW symmetry \eqref{eq:KW symmetry boson}, we have 
\begin{equation}
\begin{aligned}
    &\lim_{|\eta|\rightarrow \infty}\langle \cos {\Phi(\eta,\bar\eta)}\cos{\Phi(0)}\rangle_{\text{MR}_\infty}>0\,,\\
    &\lim_{|\eta|\rightarrow \infty}\langle \cos \frac{\Theta(\eta,\bar\eta)}{2}\cos\frac{\Theta(0)}{2}\rangle_{\text{MR}_\infty}>0\,.
\end{aligned}
\label{eq:phi phi theta theta}
\end{equation}
Thus both the $U(1)_m$ and $U(1)_w$ symmetry are broken spontaneously. 
On the other hand, $\Theta$ creates vortices of $\Phi$ (and vice versa), which are linearly confined in the symmetry breaking phase so that their mixed correlators still decay exponentially, e.g. 
\begin{equation}
\begin{aligned}
    \langle\cos{\Phi(\eta,\bar\eta)} \cos\Theta(0) \rangle_{\text{MR}_\infty} \sim \calO(e^{-|\eta|/\xi})\,.
\end{aligned}
\label{eq:phi theta}
\end{equation}
Eq.~\eqref{eq:phi phi theta theta} and \eqref{eq:phi theta} imply that a $\Phi$ condensate and a $\Theta$ condensate coexist in the reference state and they make separate contributions to the effective theory.\footnote{The $U(1)_m$ and $U(1)_w$ symmetries have a mixed anomaly. Therefore, the coexistence of $\Phi$ and $\Theta$ condensate is also a coexistence of ordered and disordered phase for each of the $U(1)$ symmetry, which typically happens at first order transitions. The current situation is even more special as the Mermin-Wagner theorem forbids continuous symmetry breaking in usual 2D theories. Our theory, not described by a local Lagrangian, can evade this no-go constraint.}
Specifically, we replace $\varepsilon_1$ and $\varepsilon_2$ by constant fields 
\begin{equation}
    \varepsilon_1\rightarrow \cos2\Phi + \cos \Theta\,,\quad \varepsilon_2\rightarrow \cos 2\Phi - \cos \Theta
\end{equation}
and neglect all correlators that mix $\Phi$ and $\Theta$.
With these recipes, we have 
\begin{equation}
\begin{aligned}
    &\quad\frac{Z_{\text{unpaired}}^{(n)}}{Z_{\text{MR}_\infty}} \\
    & \approx\frac{1}{(2n)!} \langle \Big(2\cos 2\Phi\int d^2z \cos\phi_-(z)\Big)^{2n}\rangle_{\text{MR}_\infty} \\
    & +  \frac{1}{(2n)!}\langle \Big(2 \cos \Theta\int d^2z \cos\phi_-(z)\Big)^{2n}\rangle_{\text{MR}_\infty}\,,
\end{aligned}
\end{equation}
where the two terms account for the contribution from the $\Phi$ and $\Theta$ condensate, respectively.

Altogether, the effective theory near the strong decoherence limit is
\begin{equation}
\begin{aligned}
    \tr\rho^2 \approx \int_0^{2\pi} d\Phi & \int D\phi_- e^{- S_{\text{eff}}[\phi_-, \cos2\Phi] } \\
    & + \int_0^{2\pi} d\Theta\int D\phi_- e^{- S_{\text{eff}}[\phi_-, \cos\Theta] }\,,
\end{aligned}
\label{eq:effective theory MR2}
\end{equation}
where $S_{\text{eff}}$ takes the form of a sine-Gordon theory, e.g.,
\begin{equation}
\begin{aligned}
    S_{\text{eff}} & [\phi_-, \cos2\Phi] \\
    & = \int \frac{m}{4\pi} (\nabla\phi_-)^2 -2e^{-\bar\mu}\cos 2\Phi \cos m\phi_-(x)\,.
\end{aligned}
\label{eq:effective action MR2}
\end{equation}
This closely resembles the effective description of the decohered Laughlin state in \eqnref{eq:SG}, except that the Ising sector of the MR state enters by modifying the coupling of $\cos m \phi_-$ to be proportional to $\cos 2\Phi$ or $\cos \Theta$.

Because $\Phi$ and $\Theta$ are constant fields, the perturbation retains its scaling dimension $m/2$. 
For $m < 4$, the relevance of $\cos m \phi_-$ implies no transition at any finite decoherence strength. 
Minimizing the two relevant perturbation $\cos 2 \Phi \cos m \phi_-$ and $\cos \Theta \cos m \phi_-$ pins $\Phi$ at four discrete values $0,\pi/2,\pi, 3\pi/2$, and $\Theta$ at $0,\pi$.
One can verify that $\Phi$ and $\Theta$ are pinned at the same sets of values for the pristine MR state.
Therefore, it suggests that the system remains fully topological at all finite decoherence

For $m > 4$, a transition occurs at a finite threshold, beyond which $\phi_-$ becomes a critical degrees of freedom. 
To understand the dynamics of the $\Phi$ and $\Theta$ field, we can integrate out $\phi_-$ and have
\begin{equation}
\begin{gathered}
    \tr\rho^2 = \int_0^{2\pi}d\Phi e^{-F[\cos 2\Phi]} + \int_0^{2\pi}d\Theta e^{-F[\cos \Theta]}\,, \\
    F[X] = -\log\int D\phi_-e^{- S_{\text{eff}}[\phi_-, X]}
\end{gathered}
    \label{eq:eff_MR3}
\end{equation}
For $m \geq 4$, $\cos m \phi_-$ is irrelevant or marginal so that we can perturbatively expand $F[X]$ in $e^{-\bar{\mu}}$ and have
\begin{equation}
\begin{aligned}
F[X] - F[0] \approx & \\
- 2e^{-2\bar\mu} X^2 & L^2 \int d^2 x \langle \cos m\phi_-(x)\cos m\phi_- (0)\rangle\,,
\end{aligned}
\end{equation}
where $F[0]$ is the free energy at $e^{-\bar\mu}=0$ and translation invariance simplifies the expression. 
For $m \geq 4$, the integral converges to a finite positive value that depends on the short-distance cutoff, and we have 
\begin{equation}
    F[X] - F[0] \approx - \frac{2e^{-2\bar{\mu}} L^2 a^{2 - m}}{m - 2}  X^2\,,\,
    \label{eq:FX}
\end{equation} 
where $a$ is a short-distance cutoff. Although the perturbation is not relevant, it can generate new symmetry-allowed terms that fully determines the dynamics of $\Phi$ and $\Theta$. Indeed, \eqnref{eq:eff_MR3} and \eqref{eq:FX} are the simplest form compatible with the KW$\times$ KW symmetry.
As a result, $\Phi$ and $\Theta$ are still pinned at the same sets of values for any finite decoherence.
It suggests that the Ising sector remains intact despite  the fact that the bosonic charge sector is corrupted above the threshold.
In what follows, we discuss the precise information-theoretic meaning of this transition.

\subsection{Integrity of the fusion space\label{sec:MR_anyon}}

In this section, we examine how the decoherence affects the fusion space of multiple fundamental quasi-holes. 
Specifically, we apply decoherence to the state \eqref{eq:4qhs_encoding} that maximally entangles the fusion space of four quasi-holes with a reference qubit, and evaluate the R\'enyi-2 quantum coherent information
$$
    I_c^{(2)} = \log \frac{\tr \rho_{RQ}^2}{\tr \rho_Q^2}\,.
$$
We will show that both the numerator and denominator can be mapped to simple correlation functions in the effective theory \eqref{eq:effective theory MR2}, enabling a straightforward computation of $I_c^{(2)}$ near the strong decoherence limit.

Using the quasi-hole wave functions \eqref{eq:four_qhs} and the relation between conformal blocks and correlation functions of non-chiral primaries \eqref{eq:foursigma}, we have
\[
\begin{aligned}
\tr\rho_{RQ}^2 & = \int \calD(z^1_i,z^2_j) e^{- V(\{z_i^1\}, \{z_i^2\})} \\
& \langle  \prod_{k=1}^4 e^{i\phi_+}\sigma_{12}(\eta_k,\bar \eta_k) \prod_{\alpha,i} e^{im\phi_\alpha} \varepsilon_\alpha(z^\alpha_i,\bar z^\alpha_i)\rangle_{\text{bg}}
\end{aligned}
\label{eq:RQ_MR}
\]
where $\sigma_{12} = \sigma_1\sigma_2$, $\eta_{k=1,\ldots,4}$ are locations of the four quasi-holes.
The only difference between $\tr \rho_{RQ}^2$ and the purity $\tr \rho^2$ is the insertion of 
$$
\prod_{k}^4 e^{i\phi_+}\sigma_{12}(\eta_k,\bar \eta_k)\sim  \prod_{k}^4 e^{i\phi_+}\cos \Phi(\eta_k,\bar\eta_k)\,.
$$
In the effective theory \eqref{eq:effective theory MR2}, we can ignore $\phi_+$ and approximate $\cos \Phi(\eta_k,\bar\eta_k)$ by a constant field. In the effective theory \eqref{eq:effective theory MR2}, the above 4-point correlator vanishes in the $\Theta$ condensate, and can be approximated by a constant field $\cos^4\Phi$ in the $\Phi$ condensate \eqref{eq:effective action MR2}.
We have
\begin{equation}
    \tr\rho_{RQ}^2 = \calN \langle (\cos {\Phi})^4 \rangle_\text{eff},
\end{equation}
where $\langle \ldots \rangle_\text{eff}$ stands for the expectation value in the effective theory  \eqref{eq:effective action MR2}, and $\calN = \int \calD(\Phi, \phi_-) e^{ -S_{\text{eff}}  [\phi_-, \cos2\Phi]}$ is a normalization factor.
Analogously, $\tr\rho_Q^2$ corresponds to inserting the following operator (ignoring $e^{i\phi_+}$'s): 
\begin{equation*}
\begin{aligned}
\frac{1}{2} \big( &\sigma_{12}(\eta_1,\bar \eta_1)\sigma_{12}(\eta_2,\bar\eta_{2})\sigma_{12}(\eta_3,\bar\eta_3)\sigma_{12}(\eta_4,\bar\eta_4) \\
+ & \mu_{12}(\eta_1,\bar \eta_1)\mu_{12}(\eta_2,\bar\eta_{2})\sigma_{12}(\eta_3,\bar\eta_3)\mu_{12}(\eta_4,\bar\eta_4)\\
+ & \sigma_{12}(\eta_1,\bar \eta_1)\mu_{12}(\eta_2,\bar\eta_{2})\sigma_{12}(\eta_3,\bar\eta_3)\mu_{12}(\eta_4,\bar\eta_4)\\
+ & \mu_{12}(\eta_1,\bar \eta_1)\sigma_{12}(\eta_2,\bar \eta_2)\sigma_{12}(\eta_3,\bar \eta_3)\mu_{12}(\eta_4,\bar \eta_4) \big)\,.\\
\end{aligned}
\end{equation*}
In the effective theory, we get 
\begin{equation}
    \tr\rho_{Q}^2 = \frac{\calN}{2}\langle (\cos {\Phi})^4\rangle_{\text{eff}}+\frac{3\calN}{2}\langle (\cos {\Phi}\sin \Phi)^2\rangle_{\text{eff}} \,,
    \label{eq:Q_MR}
\end{equation}
which, similarly, involves only the $\Phi$ condensate to evaluate the expectation value. Therefore, the R\'enyi-2 quantum coherent information is simplified as 
\begin{equation}
    I_c^{(2)} = \log \frac{2\langle (\cos {\Phi})^4 \rangle_\text{eff}}{\langle (\cos {\Phi})^4\rangle_{\text{eff}} + 3\langle (\cos {\Phi}\sin \Phi)^2\rangle_{\text{eff}}}
\end{equation}
in the effective theory.

We first calculate $I_c^{(2)}$ at the strong decoherence limit as a sanity check of our effective theory. In this case, $\Phi$ and $\phi_-$ are decoupled from each other, which reduces the expectation value to an integral over $\Phi$, e.g., 
\begin{equation*}
    \braket{\cos^4 \Phi}_{\text{eff}} =\frac{1}{2\pi} \int_0^{2\pi} d\Phi \cos^4 \Phi\,,\quad (e^{-\bar\mu} =  0)\,.
\end{equation*}
We thus obtain the expected result
\begin{equation}
\begin{aligned}
    I_c^{(2)} = 0\,,\quad (e^{-\bar\mu} =  0)\,.
\end{aligned}
\end{equation} 

Surprisingly, infinitesimal deviation from the strong decoherence limit restores $I^{(2)}_c$ to its initial value $\log 2$ in the thermodynamic limit. 
To see this, recall that $\Phi$ remains pinned at the same four values $0,\pi/2,\pi, 3\pi/2$ regardless whether $\cos m\phi_-$ is relevant or not. In all cases, evaluating the expectation values in \eqnref{eq:RQ_MR} and \eqref{eq:Q_MR} reduces to a discrete sum. 
We get
\begin{equation}
    I_c^{(2)} = \log 2\,,\quad (0 < e^{-\bar\mu} \leq 1)\,.
    \label{eq:Ic2 finite decoherence MR}
\end{equation}
Strictly speaking, the effective theory and its corollaries are valid near the strong decoherence limit. Since we expect $I_c^{(2)}$ to be monotonous in the decoherence strength, we extend the range of validity for \eqnref{eq:Ic2 finite decoherence MR} to include finite decoherence strength.
Therefore, although the bosonic charge sector of the topological order is corrupted across the BKT transition, the fusion space of the non-Abelian fundamental quasi-holes remain well preserved for any finite dephasing. 
In a finite-size system, the perturbation becomes negligible if $\bar{\mu} \sim \calO(\log L)$. Therefore the lifetime of this quantum memory using the fusion space is also of the order $\calO(\log L)$, as in the case of the Laughlin state above the critical filling.

\subsection{Corruption of information stored in the Abelian sector}

In contrast to the robustness of the fusion space, the filling-dependent transition does affect information stored in the Abelian sector of the MR state.
Specifically, consider the MR states on the torus~\cite{read2000paired}. 
Similar to the Laughlin states, the degenerate MR wavefunctions can be constructed from the correlation function of the chiral Ising and compact boson CFT~\cite{read2009non-Abelian}:
\begin{equation}
\begin{gathered}
    \Psi_{\alpha,\ell}(\{z_i\})=\langle \prod _i^{N_e} \psi(z_i)\rangle'_\alpha\langle\prod_i^{N_e} e^{im\varphi(z_i)}\rangle'_{\text{bg},\ell-\alpha_x+1/2}  
\end{gathered}
\end{equation}
where $\langle \ldots\rangle'$ stands for the unnormalized correlator, $\alpha=(\alpha_x ,\alpha_y)\,,\,\alpha_i=0,1/2$ labels the boundary conditions of the Majorana fermion $\psi(z+L_i)=-e^{2i\pi \alpha_i}\psi(z)$, and $\ell-\alpha_x+1/2$, $\ell = 0,1,\ldots,m-1$, labels different winding number sectors of the chiral boson CFT.
When the number of electrons $N_e$ is odd, only $\alpha=(1/2,1/2)$ gives a nonzero result and the ground-state degeneracy is $m$.
When $N_e$ is even, $\alpha=(0,0)$, $(0,1/2)$, and $(1/2,0)$ all give nonzero results, and the degeneracy is $3m$.

To address the stability of the information in the Abelian sector, we fix the Ising sector label $\alpha$ and use only $\ell$ for the encoding. 
In this case the same calculation of the coherent information as in \secref{sec:Laughlin_eff} applies here and will yield the same result. 
Namely, for systems below the critical filling $\nu = 1/4$, the information encoded with $\ell$ (characterized by R\'enyi-2 quantities) will be corrupted across the decoherence-induced transition.
The corrupted quantum information exactly corresponds to the CM motion of the fractional quantum Hall states. 
Similar to the Laughlin states on torus, $\Psi_{\alpha,\ell}(\{z_i\})$ factorizes into a Gaussian part, a holomorphic CM-motion part, and a holomorphic relative-motion part~\cite{chung2007explicit}:
\begin{equation}
    \Psi_{\alpha,\ell}(\{z_i\})=e^{-\sum_i y_i^2}F_{\alpha_x,\ell}(Z)f_\alpha(\{z_i-z_j\})\,.
\end{equation}
Importantly, changing $\ell$ only affects the CM part, and the corresponding logical operators are, again, the magnetic translation operator $T_x,T_y$. 
Therefore, we can restate the result as: the quantum memory of the CM motion is corrupted above the threshold.

In principle we can use both $\alpha$ and $\ell$ to encode quantum information on the torus.
Calculating the coherent information in this general case, especially that from the Ising sector, is beyond the scope of this work. We leave it to the future work.

\subsection{Critical filling for the von Neumann quantities\label{sec:MR_vN}}

In this section, we extend our discussion to the von Neumann quantities. Using the same approach as in \secref{sec:laughlin_vN}, we analyze the von Neumann entropy to obtain both the upper and lower bounds of the critical filling $\nu_c$. 

For the MR state, the instability of the von Neumann entropy near the strong decoherence limit also obeys the general inequality \eqref{eq:bound_DeltaS}. Again, we are interested in $\delta S_L$, the contribution of macroscopically separated pairs with $|\eta-\eta'|>\alpha L$ $(0<\alpha<1)$: 
\begin{equation*}
\begin{aligned}
    \frac{1}{2N L^2} \big(& \int_{|\eta-\eta'|>\alpha L} d^2 \eta d^2 \eta' C_e^{(1)}(\eta,\eta') \big)^2 \\
    &\quad \leq |\delta S_L| \leq \int_{|\eta-\eta'|>\alpha L} d^2 \eta d^2 \eta' C_e^{(1)}(\eta,\eta')\,,
\end{aligned}
\end{equation*}
where $C_e^{(1)}(\eta,\eta')$ is the R\'enyi-1 correlator~\eqref{eq:Renyi1_0} for the fully decohered MR state $\rho_\infty$
\begin{equation*}
    C_e^{(1)}(\eta,\eta')=\tr  \big( c(\eta')\sqrt{\rho_\infty}c^\dagger (\eta')c(\eta)\sqrt{\rho_\infty}c^\dagger(\eta) \big)\,.
\label{eq:Renyi1_MR1}
\end{equation*}
Unlike the decohered Laughlin case, the R\'enyi-1 correlator itself lacks a direct plasma analogy. Nevertheless, as shown in \appref{app:Renyi1_MR}, it admits upper and lower bounds that do map onto a plasma description. 
For separation $|\eta-\eta'|\gg \xi$ (screening length of the plasma), 
\begin{equation}
    \frac{c'}{|\eta-\eta'|^{m/2+3/4}}\leq C_e^{(1)}(\eta,\eta')\leq \frac{c}{|\eta-\eta'|^{m/2}}\,,    
\end{equation}
where $c,c'$ are some $\calO(1)$ constants.
Integrating these bounds yields
\begin{equation}
C'L^{5/2-m}\leq |\delta S_L|\leq C L^{4-m/2},
\end{equation}
for some finite constant $C,C'$. 
Therefore, the critical filling for von Neumann quantities satisfies
\begin{equation}
    1/8 \leq \nu_c(1) \leq 1/3\,.
\end{equation}
The experimentally relevant Moore-Read state at $\nu = 1/2$ lies above this range, and it therefore remains topological at any finite decoherence strength.

\section{Discussion and outlook}
\label{sec:discussions}

In this work, we examined the Laughlin and Moore-Read wave functions subject to decoherence that couples to the electron density. We identified a critical filling factor, below which the system can exhibit a BKT transition into a critical phase. Within this phase, the quantum information encoded in the degenerate ground states is partially corrupted, whereas the information stored in the fusion space of non-Abelian anyons (for the Moore-Read states) remains intact. 

These results and, in particular, the existence of a critical filling factor, derive from the properties of the fully decohered states.
Surprisingly, the fully decohered states are always critical, with a varying scaling dimension, making them unstable to infinitesimal coherence above a critical filling. For this reason, the system exhibits absolute resilience to decoherence above the critical filling.
If the fully decohered state were ``gapped", in the sense of having a finite correlation length and Markov length \cite{sang2025stability, sang2025mixed}, then it would have a finite domain of stability and the transition would occur at a finite decoherence strength regardless of filling factor ~\cite{fan2023diagnostics,bao2023mixed,lee2023quantum,wang2025intrinsic,sang2024mixed,su2024tapestry}. 

The fully decohered states are interesting and unusual in other aspects too. Having zero coherent information, rather than the minimum value of $-\log m$ implies that they retain $\log m$ bits of residual information. But although the fully decohered states are classical by virtue of having a diagonal density matrix in the position basis, the retained memory cannot be entirely classical as we explain below. 

In abelian topologically ordered states the retained memory is determined by the subset of anyons that have not been proliferated by the decoherence and that have a trivial braiding with those that have been proliferated \cite{bao2023mixed}. This memory is said to be classical if the residual nontrivial logical (anyon) operators form a commuting algebra. 
One example is the fully decohered toric code under $X$ or $Z$ errors. If the $X$ strings proliferate due to $X$ errors, the commuting $X$ loops on the two cycles persist as a classical memory. 

In the Laughlin state, on the other hand, the loop operators $T_x$ and $T_y$ transporting quasi-holes along the two cycles of the torus, form non-commuting logical operators. Because
$C_4$ rotation symmetry is preserved by decoherence, it is impossible to preserve only a commuting subset of nontrivial logical operators and thereby a purely classical memory. 
All the logical information would be destroyed if the quasi-holes fully proliferate because of their nontrivial self statistics. However, for the density dephasing noise we studied here, the quasi-holes are not fully proliferated, and instead establish a critical state, leading to zero (rather than negative) coherent information. This indicates that the logical information encoded by $T_x, T_y$ is partially preserved. 
The analysis also holds for the information encoded in the center-of-mass motion of Moore-Read states on the torus. A similar analysis also applies to the toric code under $(X+Z)$ noise \cite{chen2024unconventional}, which leads to a critical state at infinite decoherence. However, this requires a fine-tuned $e$-$m$ symmetry, and therefore the critical state exists only at a single point and not as an extended critical phase. 

Throughout this work, we have analyzed dynamics generated solely by dephasing, coupling to the microscopic electron density \eqref{eq:density dephasing}.
Considering this noise model allowed us to develop an analytical solution, and provides a guideline to investigate other, possibly more realistic, dynamics.
For example, the noise model considered here can be viewed as a large bandwidth noise, producing excitations at all energy scales. Certain systems, however, could be dominated by low frequency noise that creates excitations below the Landau level gap. Moreover, in solid state realizations of these states the decoherence operates in conjunction with the intrinsic Hamiltonian dynamics. Taken together, the Hamiltonian dynamics and the noise can give rise to incoherent creation and diffusion of anyons. The effect of this kind of phenomenological noise-model on Laughlin states will be addressed in a forthcoming publication.

It is also interesting to consider noise models that break the strong global $U(1)$ symmetry of the problem. A realistic example is the incoherent hopping of electrons in and out of the system. If this process is weak compared to the density dephasing, we can treat it in the vicinity of the fully decohered (density-dephased) critical state using the formalism developed in this work.
To illustrate the basic idea, consider the $\nu = 1/m$ Laughlin states subject to weak electron non-conserving noise.
Using the representation of the electron operator in the field theory  $c\sim e^{im\varphi}$, the effective field theory for the purity \eqref{eq:SG} is generalized to 
\begin{equation}
    S_{\text{eff}} = \int \frac{m}{4\pi} (\nabla\phi_-)^2 - 2e^{-\bar{\mu}} \cos m\phi_- -g \cos \theta_-\,,
    \label{eq:Seff particle loss}
\end{equation}
where $\bar{\mu} \gg 1$ and $g \ll 1$ are the strengths of the density dephasing and electron loss process respectively.
The two perturbations share the same scaling dimension $[\cos \theta_-]=[\cos m\phi_-]=m/2$. 
As a result, \eqnref{eq:Seff particle loss} realizes the self-dual sine-Gordon model~\cite{Lecheminant:2002va}.
At $m> 4$, both cosine terms are irrelevant, and the critical phase remains intact in the presence of particle loss. 
At $m\leq 4$, both terms are relevant. While $\cos m\phi_-$ tends to drive the system back to the topological phase, $\cos \theta_-$ leads to a trivial phase so that the memory lifetime above the critical filling is reduced to a finite value in the thermodynamic limit. The detailed interplay between density dephasing and particle loss is left for future study.

More generally, to better understand the behavior of non-abelian topologically ordered states under generic decoherence sources, it is worth considering simpler examples, such as the chiral Ising topological order, that is not complicated by a global $U(1)$ symmetry or a bosonic charge sector in the CFT. 
Finally, we note that the quantum coherent information, discussed throughout this work as an intrinsic measure of the transition is not a directly observable quantity. 
It is interesting to explore its implication on, e.g., the braiding experiments.
It is also interesting to construct a decoder and manifest the transition through decoding fidelity.
This may be easier to achieve in fractional quantum Hall states subject to phenomenological noise, which can be expressed directly in terms of anyon pair creation and hopping \cite{sala2025intrinsic}.

\section*{Acknowledgment}

We thank Jing-Yuan Chen, Margarita Davydova, Tarun Grover, Ethan Lake, Shengqi Sang, Chong Wang, Michael Zaletel for helpful discussions. R.F. thanks Yimu Bao and Meng Cheng for insightful discussions that partially inspired the current work. 
This work was supported in part by the NSF QLCI program through Grant No. OMA-2016245, by a Simons Investigator Award (E.A.)  and by the Gordon and Betty Moore Foundation Grant GBMF8688 (R.F. and S.G.).
Z.W. acknowledges the support from the Tsinghua Visiting Doctoral Students Foundation.
T.W. is supported by the U.S. Department of Energy, Office of Science, Office of Basic Energy Sciences, Materials Sciences and Engineering Division under Contract No. DE-AC02-05-CH11231 (Theory of Materials program KC2301).

\appendix

\section{Cheat sheet on the free-boson and Ising CFT}
\label{app:cheat sheet}

In this section, we collect basic facts about the free-boson CFT and the Coulomb-gas formalism for the Ising CFT. We refer the reader to Ref.~\cite{Ginsparg:1988ui,francesco2012conformal} for details. 

\subsection{Free-boson CFT on the plane}

Consider a non-chiral compact scalar boson $\phi$ with the following Euclidean action
\begin{equation}
    S[\phi] = \frac{g}{4\pi} \int (\nabla \phi)^2 d^2 z\,,\quad \phi \simeq \phi + 2\pi\,.
    \label{eq:boson action 1}
\end{equation}
The two-point function of $\phi$ (formally) reads 
\begin{equation}
    \langle\phi(z_1,\bar{z}_1) \phi(z_2,\bar{z}_2) \rangle = - \frac{\log |z_1 - z_2|}{g}\,.
\end{equation}
Another commonly used convention is to absorb the coupling constant into the radius of the boson
\begin{equation}
    S[\phi'] = \frac{1}{8\pi} \int (\nabla \phi')^2 d^2 x\,,\quad \phi' \simeq \phi' + 2\pi R\,.
    \label{eq:boson action 2}
\end{equation}
The two conventions are related by $g = R^2/2$, $\phi = \phi'/R$.
Here, we work with the first one \eqref{eq:boson action 1}.

The action manifests two $U(1)$ symmetries that are generated by the following two conserved currents
\begin{equation}
    J_\mu = \frac{1}{2\pi} \partial_\mu \phi\,,\quad \tilde{J}_\mu = \frac{1}{2\pi} \epsilon_{\mu\nu} \partial_\nu \phi\,.
\end{equation}
They generates the shift and winding of $\phi$, respectively.
The local operators, especially the vertex operators, can be organized by charges under the two symmetries.

Specifically, there are two basic types of vertex operators, the electric and magnetic operators.
One convenient way of writing them down is to separate the non-chiral field $\phi(z,\bar{z})$ into a chiral (holomorphic) part $\varphi(z)$ and anti-chiral part $\bar{\varphi}(\bar{z})$ 
\begin{equation}
    \phi(z,\bar{z}) \equiv \varphi(z) + \bar{\varphi}(\bar{z})\,,
\end{equation}
which have the following the correlation functions
\begin{equation}
\begin{aligned}
    \langle \varphi(z_1) \varphi(z_2) \rangle = -\frac{\log(z_1 - z_2)}{2g}\,,\\
    \langle \bar{\varphi}(\bar{z}_1) \bar{\varphi}(\bar{z}_2) \rangle = -\frac{\log(\bar{z}_1 - \bar{z}_2)}{2g}\,.
\end{aligned}
\end{equation}
The electric and magnetic operators are defined as (we suppress normal orderings in the expression)
\begin{equation}
\begin{aligned}
    V_e(z,\bar{z}) \equiv & e^{i e (\varphi(z)+\bar{\varphi}(\bar{z}))}\,, \\
    \calO_w(z,\bar{z}) \equiv & e^{i w g (\varphi(z) - \bar{\varphi}(\bar{z}))}\,,
\end{aligned}
\end{equation}
where one often calls $\theta = g (\varphi - \bar{\varphi})$ the dual field.
One can also define their composite, the electromagnetic operator
\begin{equation}
\begin{gathered}
    V_{e,w} = :V_e \calO_w: = :e^{i 2\sqrt{g} \alpha_{e,w} \varphi + i 2 \sqrt{g} \tilde{\alpha}_{e,w} \bar{\varphi}}:\,, \\
    \alpha_{e,w} = \frac{1}{2} \big(\frac{e}{\sqrt{g}} + \sqrt{g} w \big)\,,\,
\tilde{\alpha}_{e,w} = \frac{1}{2} \big( \frac{e}{\sqrt{g}} - \sqrt{g} w \big)\,,
\end{gathered}
\label{eq: Vew}
\end{equation}
which has the conformal dimension
\begin{equation}
h_{e,w} = \alpha_{e,w}^2 \,,\,
\tilde{h}_{e,w} = \tilde{\alpha}_{e,w}^2\,.
\end{equation}
The shift and winding symmetries imply that the correlation function of these operators is nonzero only when they satisfy the two neutrality conditions
\begin{equation}
    \sum_k e_k = \sum_a w_a = 0\,,
\end{equation}
and we have
\begin{equation}
\begin{aligned}
\langle \prod_k V_{e_k,w_k} & (z_k,\bar{z}_k) \rangle = \\
\prod_{i<j} & (z_i - z_j)^{2\alpha_i \alpha_j} (\bar{z}_i - \bar{z}_j)^{2\tilde{\alpha}_i \tilde{\alpha_j}}\,.
\end{aligned}
\label{eq:e m correlator}
\end{equation}
A charge-$e$ electric operator encircling a charge-$w$ magnetic operator produces a phase $2\pi ew$.
More generally, the chiral and anti-chiral fields $\varphi$, $\bar{\varphi}$ circling around $V_{e,m}$ are shifted by
\begin{equation}
    \varphi \mapsto \varphi + \pi (w + e/g)\,,\quad
    \bar{\varphi} \mapsto \bar{\varphi} + \pi (w - e/g)\,,
    \label{eq:chiral field boundary condition}
\end{equation}
or equivalently,
\begin{equation}
    \phi \mapsto \phi + 2\pi w\,,\quad
    \theta \mapsto \theta + 2\pi e\,.
\end{equation}
In a set of mutually local vertex operators, $e$ and $w$ must be integers.
It follows that $\theta \simeq \theta + 2\pi$.

\subsection{Coulomb-gas formalism for the Ising CFT}
\label{app:Coulomb gas}

In this section, we briefly review the Coulomb-gas formalism for the Ising CFT. In particular, we show how to represent the thermal operator $\varepsilon$ and spin operator $\sigma$ in this formalism.
There are several ways to present this formalism. Here we will not be pedantic and just pick one presentation that can give us the correct answer. We refer interested readers to Chapter 9 of \cite{francesco2012conformal} and references therein for a more thorough discussion.

We start with the linear dilaton CFT, which has the same action \eqnref{eq:boson action 1} but is defined for a non-compact boson and a different energy-momentum tensor.
To emphasize the difference, we denote the boson field by $X$ and the chiral energy-momentum tensor reads
\begin{equation}
    T(z)=-\frac{1}{2}:\partial X\partial X:+i\sqrt{2}\alpha_0\partial^2 X\,,
\end{equation}
where $\alpha_0$ parameterizes the CFT.
It follows from the $TT$ OPE that the central charge is $c=1-24\alpha_0^2$.
The vertex operators $V_\alpha =\,:\!e^{i\sqrt{2}\alpha X}\!:$ are the scalar primaries, where the non-compactness of $X$ allows $\alpha$ to be any real number. 
From the OPE
\begin{equation}
    T(z)V_\alpha(0)\sim \frac{\alpha^2-2\alpha\alpha_0}{z^2} V_\alpha (0)+\frac{\partial V_\alpha(0)}{z}\,,
\end{equation}
we can read out its conformal dimension $h_\alpha=\alpha^2-2\alpha\alpha_0$.

The Coulomb-gas formalism is a prescription to reproduce all the minimal models from the linear dilaton CFT by choosing $\alpha_0$ to match the central charge and appropriate set of vertex operators to match the operator content.
To obtain the Ising CFT, we first set $\alpha_0=\sqrt{3}/12$ to get the correct central charge $c=1/2$. 
Then we select the set of vertex operators with the following $\alpha$'s so that their conformal dimensions match those of primary fields in the Ising CFT (the choice is not unique): 
\begin{equation}
    \alpha_\bbI=0\,, \quad \alpha_\varepsilon=\sqrt{3}/{2}\,,\quad\alpha_\sigma=\sqrt{3}/4\,.
\end{equation}
However, directly identifying the associated non-chiral vertex operators $V_\alpha (z,\bar z)$ with the Ising primaries leads to a contradiction. 
For example, the two-point function of the thermal operator becomes 
$\langle \varepsilon(z,\bar{z}) \varepsilon(0) \rangle = \langle V_{\alpha_\varepsilon} (z,\bar{z}) V_{\alpha_\varepsilon}(0)\rangle = 0$, where the first equality is the naive identification and second equality is the charge neutrality condition, i.e. $\langle \prod _i V_{\alpha_i}(z_i,\bar z_i) \rangle$ is nonzero if and only if $\sum_i \alpha_i=0$.

The key idea of the Coulomb-gas formalism is to remedy this issue by introducing two screening operators
\begin{equation}
    Q_\pm = \oint dw \oint d\bar{w} e^{i\sqrt{2}\alpha_\pm X(w,\bar{w})}\,.
\end{equation}
They carry nonzero charges $\alpha_\pm$ but a zero conformal dimension, i.e., $h_{\alpha_\pm}=\alpha_\pm^2-2\alpha_\pm \alpha_0=1$. 
For the Ising CFT, we have $\alpha_+ =2/\sqrt{3},\alpha_-=-\sqrt{3}/2$.
Consequently inserting these screening operators into correlation functions of vertex operators can screen their charges without affecting their conformal properties. 
Specifically, we define the screened vertex operator
\begin{equation}
V^{rs}_\alpha(z,\bar z) =  Q_-^r Q_+^s V_\alpha (z,\bar z)\,,
\end{equation}
where the integration contours in the screening operators circles the vertex operator. These contours are implicitly radially ordered and can be freely deformed as long as it does not pass other operators.
We can instead use appropriate screened vertex operators so that the total charge is neutralized and the result becomes nonzero.
We identify the primaries in the Ising CFT with these screened vertex operators, where the screening charges $r,s$ is chosen so that the correlation function obeys the charge neutrality.

For example, the correlation function of an even number of thermal operators $\varepsilon$'s can be written as
\begin{equation}
\begin{aligned}
    \langle \varepsilon (z_1, & \bar z_1) \varepsilon(z_2,\bar z_2) \rangle = \langle V_{\alpha_\varepsilon}^{20} (z_{1},\bar{z}_{1})V^{00}_{\alpha_\varepsilon}(z_{2},\bar z_{2})\rangle \\
    =& \oint dw_{1} d\bar w_{1} dw_{2} d\bar w_{2} \frac{|w_1-w_2|^3 |z_1-z_2|^3}{\prod_{i,j}^2 |z_i-w_j|^{3}} \,,
\end{aligned}
\end{equation}
where the integration contours encloses all $z_i$'s.
Similarly, the $\sigma$ two-point function is 
\begin{equation}
\begin{aligned}
    \langle\sigma(\eta_1,\bar\eta_1) & \sigma(\eta_2,\bar\eta_2)\rangle = \langle V^{00}_{\alpha_\sigma}(\eta_1,\bar \eta_1)V^{10}_{\alpha_\sigma}(\eta_2,\bar\eta_2)\rangle \\
    = & \oint dw d\bar{w} \frac{|\eta_1-\eta_2|^{3/4}}{|\eta_1-w|^{3/2} |\eta_2-w|^{3/2}}\,.
\end{aligned}
\end{equation}
We can also combine and generalize the above results to write down the correlation function of two spin operators and $2N$ thermal operators
\begin{equation}
\begin{aligned}
&\langle \sigma(\eta_1,\bar\eta_1)\sigma(\eta_2,\bar\eta_2) \prod_{i=1}^{2N}\varepsilon(z_i,\bar z_i) \rangle\\
=& \oint \prod_{k=1}^{2N+1} dw_k d\bar{w}_k |\eta_1-\eta_2|^{3/4} \prod_{i<j}^{2N} |z_i-z_j|^3 \\
& \prod_{k<l}^{2N+1}|w_k-w_l|^3\prod_{i=1}^{2N}\prod_{k=1}^{2N+1}|z_i-w_k|^{-3}\\
&\prod_{k=1}^{2N+1} |\eta_1-w_k|^{-3/2} |\eta_2-w_k|^{-3/2}  \\
&\prod_{i=1}^{2N}|\eta_1-z_i|^{3/2} |\eta_2-z_i|^{3/2}
\end{aligned}
\label{eq:coulomb gas two sigmas}
\end{equation}
As a final remark, we can apply the so-called Mathur's trick to further replace each pair of the contour integrals in the above expressions by a 2D integral over the complex plane~\cite{MATHUR1992quantum} 
\begin{equation}
    \oint dw \oint d\bar w \mapsto \int d^2 w\,.
\end{equation}
The resulting expressions are used to derive the plasma analogy for the Moore-Read state\cite{bonderson2011Plasma}.

\section{Critical filling at general R\'enyi indices and its replica limit}
\label{app:replica_limit}

The goals of this Appendix are two-fold. We will derive the R\'enyi-$n$ critical filling \eqref{eq:critical filling for n} at $n = 2,3,\ldots$ for the decohered Laughlin state. 
Then we show how to take the replica limit $n\rightarrow 1$ correctly, which complements the direct analysis of von Neumann quantities in \secref{sec:laughlin_vN}.

The basic strategy is to perform a strong-decoherence expansion of the R\'enyi-$n$ entropy 
\begin{equation}
    S^{(n)}[\rho]=\frac{1}{1-n}\log \tr \rho^n\,,
\end{equation}
and examine the instability.
Specifically, we expand the density matrix as in Eq.~\eqref{eq:large_mu_expansion}: 
$$
    \rho= \rho_\infty+y\delta \rho+y^2\delta\rho^{(2)} +O(y^3)\,,
$$
where $y=e^{-\bar\mu}$, $\rho_\infty$ and $\delta\rho$ are defined in \eqnref{eq:large_mu_expansion}, and $\delta\rho^{(2)}$ denote terms with two particles having different bra/ket positions.
Accordingly the expansion of the R\'enyi-$n$ entropy with respect to $y$ reads
\begin{equation}
    S^{(n)}[\rho] =S^{(n)}[\rho_\infty] + \delta S^{(n)}[\delta\rho] + \calO(y^3)\,,
\end{equation}
where the $\calO(y)$ term vanishes and $\delta\rho^{(2)}$ does not enter at this order. We isolate contributions to $\delta S^{(n)}$ that come from far separated unpaired particles by defining 
\begin{equation}
    \delta S^{(n)}_L=\delta S^{(n)}[\delta\rho']\,,
\end{equation}
where $\delta\rho' = \sum_{[b,c]} \psi_b\psi_c^*|b\rangle\langle c|$ includes pairs of configurations described below
\begin{equation*}
    \Bigl\{\, b=\{z_i,\eta\},\; c=\{z_i,\eta'\} \, \big | \, \,  |\eta-\eta'|>\alpha L \,\Bigr\},
\end{equation*}
with $0<\alpha<1$ and $L$ the linear system size.
We determine the critical filling $\nu_c$ by examining whether $\delta S^{(n)}_L$ diverges in the thermodynamic limit $L\rightarrow+\infty$.

\subsection{Higher integer R\'enyi indices}

For $n= 2,3,4,\cdots$, $\tr \rho^n$ has a polynomial expansion in $\delta \rho$, which yields
\begin{equation}
\delta S^{(n)}_L=\frac{ny^2}{2(1-n)}\sum_{k=0}^{n-2}\frac{\tr\rho_\infty^{n-k-2}\delta\rho'\rho_\infty^k\delta\rho'}{\tr\rho_\infty^n}\,.
\label{eq:Renyi_integer_n}
\end{equation}
The denominator of the summand can be interpreted as a single-component plasma in the screening phase (for $n < 70/m$). Similarly, the numerator is related to the same type of plasma with two test charges. 
We have
\begin{equation}
\begin{aligned}
    &\frac{\tr\rho_\infty^{n-k-2}\delta\rho'\rho_\infty^k\delta\rho'}{\tr\rho_\infty^n} = \\
    &\int_{|\eta-\eta'|>\alpha L} \frac{d^2\eta d^2\eta'}{|\eta-\eta'|^{qq'}} \int \calD z e^{-\Phi_L(z_i,\sqrt{2mn}; \eta,q;\eta',q')},
\end{aligned}
\end{equation}
where $q=\sqrt{\frac{2m}{n}}(n-k-1)$, $q'=\sqrt{\frac{2m}{n}}(k+1)$ are the charges carried by the test particles, respectively. 
Using the screening property, we have
\begin{equation}
\begin{aligned}
    \frac{\tr\rho_\infty^{n-k-2}\delta\rho'\rho_\infty^k\delta\rho'}{\tr\rho_\infty^n}
    \sim & \int_{|\eta-\eta'|>\alpha L} \frac{d^2\eta d^2\eta'}{|\eta-\eta'|^{qq'}} \\
    \sim & L^{4-\frac{2m(k+1)(n-k-1)}{n}}\,.
\end{aligned}
\end{equation}
In the thermodynamic limit, the summation in \eqref{eq:Renyi_integer_n} is dominated by the $k=0$ and $k=n-2$ term. We have
\begin{equation}
    \delta S^{(n)}_L\propto L^{4-2m(n-1)/n}\,. 
\label{eq:deltaS_integer_n}
\end{equation}
\eqnref{eq:deltaS_integer_n} diverges with the system size $L$ when the filling factor is below the following critical value
\begin{equation}
    \nu_c(n)=\frac{n-1}{2n}\,,\quad n=2,3,4,\cdots
\label{eq:nuc_integer}
\end{equation}

\subsection{Non-integer R\'enyi indices}

Next we turn to the generic case including non-integer $n$. We will show that the result \eqref{eq:nuc_integer} no longer holds for $1<n<2$. In the general case, we have to use the matrix derivative to expand $\tr \rho^n$ with respect to $\delta \rho$.
Given a diagonalizable matrix $A(y)$ and diagonal $A(0)_{ij}=\lambda_i\delta_{ij}$, the derivative of an arbitrary matrix function $f(A(y))$ is~\cite{horn1994topics}
\begin{equation}
\begin{aligned}
\frac{df(A(y))}{dy}\Big|_{y=0} &= \Delta f(\lambda_i,\lambda_j)A'(0)_{ij},\\
\frac{d^2 f(A(y))}{dy^2}\Big|_{y=0} &= 2\Delta^2 f(\lambda_i,\lambda_j,\lambda_k)A'(0)_{ik}A'(0)_{kj}\\
&~+\Delta f(\lambda_i,\lambda_j)A''(0)_{ij}\,,
\end{aligned}
\end{equation}
where we sum over repeated indices and
\begin{equation}
\begin{gathered}
\Delta f(\lambda_1,\lambda_2) = \frac{f(\lambda_1)-f(\lambda_2)}{\lambda_1-\lambda_2}\,,\,\Delta f(\lambda,\lambda)=f'(\lambda)\,,\\
\Delta^2 f(\lambda_1,\lambda_2,\lambda_3) = \frac{\Delta f(\lambda_1,\lambda_2)-\Delta f(\lambda_1,\lambda_3)}{\lambda_1-\lambda_3}\,,
\end{gathered}
\end{equation}
are the divided differences. In our case where $f(\rho)=\rho^n$, the expansion around $y=0$ is
\begin{equation}
\begin{gathered}
\tr \rho^n=\tr\rho_\infty^n+\Delta f(\lambda_a,\lambda_a)(y\delta\rho_{aa}+\frac{y^2}{2}\delta\rho^{(2)}_{aa})\\
+y^2\Delta^2f(\lambda_b,\lambda_b,\lambda_c)\delta\rho_{bc}\delta\rho_{cb}
\end{gathered}
\end{equation}
where $\lambda_a=|\psi_a|^2$. The second term vanishes, and the third term can be simplified by symmetrizing $b,c$. We have
\begin{equation}
\begin{gathered}
    \tr \rho^n=\tr\rho_\infty^n+\frac{ny^2}{2}\sum_{(b,c)}\frac{\lambda_b^{n-1}-\lambda_c^{n-1}}{\lambda_b-\lambda_c}\lambda_b\lambda_c\,,
\end{gathered}
\end{equation}
leading to
\begin{equation}
\delta S^{(n)}_L=\frac{ny^2}{2-2n}\frac{\sum_{[bc]}\lambda_b\lambda_c\frac{\lambda_b^{n-1}-\lambda_c^{n-1}}{\lambda_b-\lambda_c}} {\sum_a\lambda_a^n}\,.
\label{eq:Renyi_non_integer}
\end{equation}
For $n=2,3,4,\cdots$, we can factorize the numerator and reduce it to Eq.~\eqref{eq:Renyi_integer_n}. 
For non-integer $n$, however, the numerator of Eq.~\eqref{eq:Renyi_non_integer} has to be analyzed by different method. In particular, it does not have a plasma analogy. 
Instead, we rewrite the numerator as
\begin{equation*}
\begin{aligned}
    &\frac{\lambda_b\lambda_c}{n-1}\frac{\lambda_b^{n-1}-\lambda_c^{n-1}}{\lambda_b-\lambda_c}\\
    &\quad = (\lambda_b\lambda_c)^{n/2} \frac{(\lambda_b/\lambda_c)^{\frac{n-1}{2}}-(\lambda_b/\lambda_c)^{-\frac{n-1}{2}}}{(n-1)((\lambda_b/\lambda_c)^{\frac{1}{2}}-(\lambda_b/\lambda_c)^{-\frac{1}{2}})}
\end{aligned}
\end{equation*}
Noticing that the second factor monotonically increases with $n$ for $n>0$, we can derive the following inequality
\begin{equation}
\frac{\lambda_b\lambda_c}{n-1}\frac{\lambda_b^{n-1}-\lambda_c^{n-1}}{\lambda_b-\lambda_c}
\leq (\lambda_b\lambda_c)^{n/2}\,,\quad 1<n<2\,.
\end{equation}
We can then apply plasma analogy to the upper bound and obtain
\begin{equation}
|\delta S_L ^{(n)}|\leq cL^{4-mn/2}\Rightarrow \nu_c(n)\geq \frac{n}{8}\quad(n<2)
\label{eq:mc_non_integer}
\end{equation}
where $c>0$ some finite constant. 

Extending Eq.~\eqref{eq:deltaS_integer_n} to $n<2$ clearly contradicts the bound derived above. This implies that the critical filling $\nu_c(n)$ is not an analytic function of the R\'enyi index $n$ on the half plane $\text{Re} (n)>1$.
The origin of this non-analyticity is that $\nu_c$ is defined only in the thermodynamic limit $L\to\infty$, i.e., $\nu_c$ is extracted from the leading asymptotic behavior of $\delta S^{(n)}_L$ after discarding all subleading (finite-size) terms. 
If we instead compute the exact finite-$L$ quantity $\delta S^{(n)}$, perform the analytic continuation in $n$ at finite $L$, and only then take $L\to\infty$, we would be able to obtain the correct $\nu_c$. 
Indeed, taking the limit $n\rightarrow 1$ in the exact result~\eqref{eq:Renyi_non_integer}, we obtain
\begin{equation}
\delta S_L=-\frac{y^2}{2}\frac{\sum_{[bc]}\frac{\lambda_b\lambda_c}{\lambda_b-\lambda_c}\log \frac{\lambda_b}{\lambda_c}} {\sum_a\lambda_a},
\label{eq:deltaS_n=1}
\end{equation}
which is consistent with the result \eqref{eq:deltaS_vN} from direct perturbation expansion of the von Neumann entropy.
In other words, the analytic continuation in $n$ and taking the thermodynamic limit do not commute, so naive extrapolation from integer $n$ could fail.

\section{Perturbative calculation of quantum coherent information in the critical phase}
\label{app:perturbation}

For Laughlin states above the critical filling, the system enters a critical phase once the decoherence exceeds a threshold. In this appendix, we examine the R\'enyi-2 quantum coherent information $I_c^{(2)}$ in this critical phase.
Because the perturbation in the corresponding sine-Gordon effective theory is irrelevant, a controlled perturbative expansion applies.
We will show that the quantum coherent information remains finite throughout the critical phase and vanishes only at the strong decoherence limit.

Recall that the R\'enyi-2 quantities can be captured by the following effective Hamiltonian 
\begin{equation*}
    H_-=\int_0^{L_x} \frac{(\partial_x\theta_-)^2}{4\pi m} + \frac{m}{4\pi} (\partial_x\phi_-)^2 - 2e^{-\bar\mu}\cos m\phi_-\,.
\end{equation*}
The R\'enyi-2 quantum coherent information reads
\begin{equation*}
\begin{aligned}
    I_c^{(2)}=\log \frac{\sum_{\ell=0}^{m-1} Z_-(0,\ell)}{\sum_{\ell=0}^{m-1} Z_-(\ell,0)}\,,
\end{aligned}
\end{equation*}
where $Z_-(\ell_1, \ell_2) \approx \tr_{\ell_1, \ell_2} e^{-L_y H_-}$ is the partition function of twisted boundary conditions for $\phi_-$ and $\theta_-$ as is specified in \eqnref{eq:w_+=0}.
In the following, we assume $L_x = L_y = L$ and suppress the subscript ``$-$" on $H, \phi,\theta$ for notational simplicity.

The cosine term in the effective sine-Gordon field theory is irrelevant, and can be treated as a perturbation.
Up to the order $\calO(e^{-2\bar\mu})$, the free energy of the twisted sector $(\ell_1,\ell_2)$ changes by
\begin{equation}
\begin{aligned}
F_{(\ell_1,\ell_2)} & \approx F_{(\ell_1,\ell_2)}^{(0)} \\
- & 2e^{-2\bar\mu} \int d^2\mathbf{r} d^2\mathbf{r}' \langle e^{im\phi(\mathbf{r})} e^{-im\phi(\mathbf{r}')}\rangle_{(\ell_1,\ell_2)}\,,
\end{aligned}
\label{eq:delta F expansion}
\end{equation}
where $\langle \ldots \rangle_{(\ell_1,\ell_2)}$ denotes the expectation value in the free-boson theory with the specified boundary condition. 
We separate $\phi$ and $\theta$ into the quantum and classical part $\phi(\mathbf{r}) = \phi_{\text{qu}}(\mathbf{r}) + 2\pi w x /L$, $\theta(\mathbf{r}) = \theta_{\text{qu}}(\mathbf{r}) + 2\pi e x /L$, where $w\in \ell_1/m + \bbZ$, $e/m \in \ell_2/m + \bbZ$ in the given twisted sector $(\ell_1,\ell_2)$. Namely, the quantum part is periodic in space and the classical part accounts for the winding. 
Accordingly, the two-point function separates into a classical and quantum part as well
\begin{equation}
\begin{aligned}
    & \langle e^{im\phi(\mathbf{r})} e^{-im\phi(\mathbf{r}')}\rangle_{(\ell_1,\ell_2)} \\
    = & \sum_{(w,e)} p^{(w,e)}_{(\ell_1,\ell_2)} e^{ \frac{2\pi i m w}{L} (x - x')} \langle e^{im \phi_{\text{qu}}(\mathbf{r})} e^{-im\phi_{\text{qu}}(0)} \rangle_{\text{qu}}\,,
\end{aligned}
\end{equation}
where the summation is restricted to the allowed values in the twisted sector and $p^{(w,e)}_{(\ell_1,\ell_2)}$ is the Boltzmann weight (probability) for the boundary condition, i.e.,
\begin{equation}
    p^{(w,e)}_{(\ell_1,\ell_2)} = \frac{e^{-\pi m (w^2 + (e/m)^2)}}{\sum_{(w',e')} e^{-\pi m (w'^2 + (e'/m)^2)}}\,.
\end{equation}
The expectation value $\langle \ldots \rangle_{\text{qu}}$ in the ``quantum" part has an exact expression in terms the elliptic theta function
\begin{equation}
    \langle e^{im \phi_{\text{qu}}(\mathbf{r})} e^{-im\phi_{\text{qu}}(0)} \rangle_{\text{qu}} = \left(\frac{a}{L}\right)^m \Big| \frac{\partial_z \theta_1(0|\tau)}{\theta_1(z/L|\tau)} \Big|^{m}\,.
\end{equation}
where $a \ll L$ is a short-distance cutoff.
Plugging these results into \eqnref{eq:delta F expansion} and writing the integral in a dimensionless form, we have 
\begin{equation*}
\begin{aligned}
    F_{(\ell_1,\ell_2)} & \approx F_{(\ell_1,\ell_2)}^{(0)} - \frac{2e^{-2\bar\mu} a^m}{ L^{m-4}} \\
    \times  \sum_{(w,e)} & p_{(w,e)} \int_{[0,1]^2} d^2 \mathbf{r}\Theta(|\mathbf{r}|-\frac{a}{L}) e^{2\pi i mw x} \Big| \frac{\partial_z \theta_1(0|\tau)}{\theta_1(z|\tau)} \Big|^{m}\,,
\end{aligned}
\end{equation*}
where $\Theta(\cdot)$ is the Heaviside step function, imposing the short-distance cutoff.
For $m > 4$, the integral is dominated by the ultra-violet divergence
\begin{equation*}
\begin{aligned}
    & \int_{[0,1]^2} d^2 \mathbf{r} \Theta(|\mathbf{r}|-\frac{a}{L})e^{2\pi i mw x} \Big| \frac{\partial_z \theta_1(0|\tau)}{\theta_1(z|\tau)} \Big|^{m} \\\sim & \int_{|\mathbf{r}|>\frac{a}{L}} d^2 \mathbf{r} \frac{e^{2\pi i mw x}}{|\mathbf{r}|^m} \\
    \sim & \int_{|\mathbf{r}|>\frac{a}{L}} d^2 \mathbf{r} \frac{1}{|\mathbf{r}|^m} \big( 1 - \frac{1}{2} \left(2\pi mw x\right)^2 \big) \\
    \sim & \frac{2\pi}{m-2} \left( \frac{a}{L} \right)^{2-m}  \Big(1 - \frac{\pi^2 m^2 (m-2) w^2}{m-4} \left( \frac{a}{L} \right)^2 \Big)
\end{aligned}
\end{equation*}
The free energy of the twisted sector reads
\begin{equation}
\begin{aligned}
    F_{(\ell_1,\ell_2)} & \approx F_{(\ell_1,\ell_2)}^{(0)} - \frac{4\pi a^2e^{-2\bar\mu} }{m-2} L^2 \\
     \times \Big( & 1 - \frac{\pi^2 m^2 (m-2)}{m-4} \left( \frac{a}{L} \right)^2 \sum_{(w,e)} p^{(w,e)}_{(\ell_1,\ell_2)} w^2 \Big)\,.
\end{aligned}
\end{equation}
There is a correction to the free-energy density that is independent of the winding, and a subleading correction depending on the winding.

To perturbatively calculate the R\'enyi-2 quantum coherent information, it is convenient to divide both the numerator and denominator by $Z(0,0)$ and recast the expression in terms of the free-energy difference between different twisted sectors 
\begin{equation*}
    I_c^{(2)}=\log \frac{1+\sum_{\ell=1}^{m-1} e^{-\Delta F_{(0,\ell)}}}{1+\sum_{\ell=1}^{m-1}e^{-\Delta F_{(\ell,0)}}}\,.
\end{equation*}
The numerator is not affected at the order $\calO(e^{-2\bar{\mu}})$, i.e. $\Delta F_{(0,\ell)} = \Delta F_{(0,\ell)}^{(0)}$.
The denominator receives a constant correction
\begin{equation}
\begin{aligned}
    \Delta & F_{(\ell,0)} \approx \Delta F_{(\ell,0)}^{(0)}  \\
    & + \frac{4\pi^3 a^4e^{-2\bar\mu}  m^2 }{m-4} \Big( \sum_{(w,e)} p^{(w,e)}_{(\ell,0)} w^2 - \sum_{(w,e)} p^{(w,e)}_{(0,0)} w^2 \Big) 
\end{aligned}
\end{equation}
Noticing that $F_{(0,\ell)}^{(0)}  = F_{(\ell,0)}^{(0)}$ and $\Delta F_{\ell,0} > \Delta F_{(0,\ell)}$, we show that and $I_c^{(2)}$ is always positive for any finite decoherence. Near the strong decoherence limit, $I_c^{(2)} \propto e^{-2\bar{\mu}}$, i.e., it vanishes exponentially.

\section{Plasma analogy of the decohered Moore-Read state}
\label{app:plasma MR state}

Similar to the Laughlin wave function, the Moore-Read (MR) wave function and the quasi-hole wave functions also have a plasma analogy~\cite{bonderson2011Plasma}. In this section, we briefly review the analogy and discuss subtleties of its proper statistical mechanical interpretation that is not explicitly addressed in Ref.~\cite{bonderson2011Plasma}. 
We then apply it to the fully decohered MR state and determine its screening properties via Monte-Carlo simulation. 

\subsection{Plasma analogy of the Moore-Read state}

Recall the definition of the un-normalized MR wave function of $N$ electrons
\begin{equation*}
    \Psi(\{z_i\}) = \langle \prod_{i=1}^N\psi(z_i)e^{im\varphi(z_i)}\rangle_{\text{bg}} \,.
\end{equation*}
Using its explicit expression, its norm square reads
\begin{equation}
\begin{aligned}
    \langle\Psi|\Psi\rangle =& \int \calD z\, \langle \prod_{i=1}^N\varepsilon(z_i, \bar{z}_i) e^{im\phi(z_i)} \rangle_{\text{bg}} \\
    =& \int Dz\, |\text{Pf}(\frac{1}{z_i-z_j})|^2\times e^{-\Phi_L(z_i,\sqrt{2m})}\,,
\end{aligned}
\end{equation}
where $\Phi_L$ is defined in Eq.~\eqref{eq:laughlin plasma}.  
As is shown in Ref.~\cite{bonderson2011Plasma}, we can combine the Coulomb-gas formalism (reviewed in \appref{app:Coulomb gas}) of the Ising CFT and the Mathur's trick to rewrite the norm square of the Pfaffian as a partition function of a two-component Coulomb gas
\begin{equation}
\begin{aligned}
& |\text{Pf}(\frac{1}{z_i-z_j})|^2 =\int\prod_{k=1}^N d^2w_k e^{-\Phi_{I}(z_i,w_k)}\,,\\
& \Phi_I(z_i,w_k) = -3\sum_{i<j}^N \log|z_i-z_j|+\log |w_i-w_j|\\
&\qquad\qquad\qquad  +3 \sum_{i,j}^N \log|z_i-w_j|\,.
\end{aligned}
\label{eq:Ising_plasma}
\end{equation}
The integral on the right-hand side has an ultraviolet divergence as $z_i\rightarrow w_j$ for any $z_i$ and $w_j$. We will comment on this subtlety the end of this section.
Altogether, we can write the norm square of the MR wave function as
\begin{equation}
\begin{gathered}
\langle\Psi|\Psi\rangle = \int \calD z \prod_{k=1}^Nd^2w_ke^{-\Phi_{L}(z_i,\sqrt{2m})-\Phi_I(z_i,w_k)}\,.
\end{gathered}
\label{eq:Z_MR}
\end{equation} 
We call $\Phi_L$ and $\Phi_I$ the energy of the Laughlin and Ising plasma, respectively.
The combined plasma has two species of mobile particles: the $N$ electrons are mapped to $N$ particles at $z_1, z_2,\cdots ,z_N$ carrying both Laughlin and Ising charges $(Q_L,Q_I)=(+\sqrt{2m},+\sqrt{3})$. 
There are also $N$ auxiliary particles at $w_1,\cdots,w_N$ carrying only negative Ising charges $(Q_L,Q_I)=(0,-\sqrt{3})$. 
For $m<70$, the Laughlin plasma and Ising plasma, when not coupled to each other, are both in the screening phase individually.
Therefore the combined plasma is in the screening phase as well~\cite{bonderson2011Plasma}.

Similarly, we can derive a plasma analogy for the quasi-hole wave function. For our purpose, it suffices to consider the case with two quasi-holes
\begin{equation}
\begin{aligned}
\Psi_{\{\eta_1,\eta_2\}} & (\{z_i\}) \\
=& \langle\prod_{k=1}^2 \sigma(\eta_k) e^{i\varphi(\eta_k)/2} \prod_{i=1}^N \psi(z_i) e^{im\varphi(z_i)}\rangle_{\text{bg}}\,.
\end{aligned}
\end{equation}
Its norm square reads
\begin{equation}
\begin{aligned}
    & \langle \Psi_{\{\eta_1,\eta_2\}} |  \Psi_{\{\eta_1,\eta_2\}} \rangle \\
    =& \int \calD z \langle \sigma(\eta_1,\bar \eta_1)e^{i\phi(\eta_1,\bar\eta_1)/2}\sigma(\eta_2,\bar \eta_2)e^{i\phi(\eta_2,\bar\eta_2)/2} \\
    &\qquad\prod_{i=1}^N \varepsilon(z_i,\bar{z}_i) e^{im\phi(z_i)} \rangle_{\text{bg}}
\end{aligned} 
\label{eq:appD5}
\end{equation}
We can use the Coulomb-gas representation \eqref{eq:coulomb gas two sigmas} and Mathur's trick to rewrite it as
\begin{equation}
\begin{aligned}
& \langle \Psi_{\{\eta_1,\eta_2\}} |  \Psi_{\{\eta_1,\eta_2\}} \rangle =\int \calD z \prod_{k=1}^{N+1} d^2w_k\\
&e^{-\Phi_L(z_i,\sqrt{2m};\{\eta_1,\eta_2\},1/\sqrt{2m})-\Phi_{I}(z_i,w_k;\{\eta_1,\eta_2\},\frac{\sqrt{3}}{2})} 
\end{aligned}
\label{eq:appD6}
\end{equation}
where $\Phi_{I}(z_i,w_k;\{\eta_1,\eta_2\},\sqrt{3}/2)$ is the energy of the Ising plasma with two test charges $(0,+\sqrt{3}/2)$ at $\eta_1,\eta_2$, i.e.
\begin{equation}
\begin{aligned}
    \Phi_{I}(z_i, & w_k; \{\eta_1,\eta_2\}, \frac{\sqrt{3}}{2}) = \Phi_{I}(z_i,w_k) - \frac{3}{4}\log|\eta_1-\eta_2| \\
    + & \frac{3}{2}\sum_{\nu=1}^2\big(\sum_{i=1}^N\log|z_i-\eta_\nu|-\sum_{k=1}^{N+1}\log|w_k-\eta_\nu|\big)
\end{aligned}
\end{equation}
Therefore the norm square of the two-quasi-hole wave function $\langle \Psi_{\{\eta_1,\eta_2\}} |  \Psi_{\{\eta_1,\eta_2\}} \rangle$ is related to the free energy cost of inserting such two test charges. By the screening property of the combined plasma, the norm square should asymptotes to a finite constant for $|\eta_1-\eta_2|\rightarrow\infty$.

We close this section with a remark on the ultraviolet divergence in \eqnref{eq:Ising_plasma}.
To cure the divergence, the most straightforward approach is to impose a short-distance cutoff: only configurations satisfying $|z_i - w_j| > r_0$ for all $i,j$ are included in the integration domain. 
The advantage of this approach, compared with analytical continuation, is that the regulated integral has a simple stat-mech interpretation: a two-component plasma with a hard-core repulsion.
However, it immediately raises a subtlety: \eqnref{eq:Ising_plasma} is no longer a mathematical equality. The left-hand side of \eqnref{eq:Ising_plasma} clearly does not depend on the short-distance cutoff $r_0$ whereas the regulated integral does. Instead, it can be shown that the regulated integral, despite of having complicated dependence on $r_0$, has a $r_0$-independent part that is equal to the Pfaffian norm square, in addition to a ultra-violet divergent part. Since the infrared properties of the plasma should be insensitive to the ultra-violet divergent part, we will directly use the right-hand side of \eqnref{eq:Ising_plasma} with a finite $r_0$ to deduce properties of the MR wave function.

\subsection{Plasma analogy of the fully decohered Moore-Read state}

It is straightforward to generalize the above idea and write down a plasma analogy for the purity of the fully decohered MR state 
\[
\begin{aligned}
 \tr\rho_\infty^2\equiv Z_{\text{MR}_\infty} & = \int \calD \xi \prod_{k=1}^N d^2w^1_k \prod_{l=1}^Nd^2w^2_l \\
&e^{-\Phi_L(\xi_p,2\sqrt{m})-\Phi_{I}(\xi_p,w^1_k)-\Phi_{I}(\xi_p,w^2_l)}\,.
\end{aligned}
\label{eq:plasma_MR_fully}
\]
Accordingly, 
\begin{equation}
\begin{aligned}
\langle \sigma_1 \sigma_2 & (\eta,\bar\eta)\sigma_1\sigma_2(0)\rangle_{\text{MR}_\infty}\\
= & \frac{1}{Z_{\text{MR}_\infty}} \int \calD \xi \prod_{k=1}^{N+1} d^2w^1_k \prod_{l=1}^{N+1}d^2w^2_l \\
&e^{-\Phi_L(\xi_p,2\sqrt{m})-\sum_{\alpha=1,2}\Phi_{I}(\xi_p,w^\alpha_k;\{\eta,0\},\frac{\sqrt{3}}{2})}\,,
\end{aligned}
\label{eq:sigma12 fully decohered}
\end{equation}
which is related to the free-energy cost of inserting test charges into the Ising plasmas.
From the bosonization dictionary \eqref{eq:dictionary} and the $\text{KW}\times \text{KW}$ symmetry \eqref{eq:KW symmetry boson}, \eqnref{eq:sigma12 fully decohered} is also proportional to $\langle\cos \Phi(\eta,\bar\eta)\cos \Phi(0)\rangle_{\text{MR}_\infty}=\langle \cos\Theta(\eta,\bar\eta)\cos\Theta(0)\rangle_{\text{MR}_\infty}$. 
In the following, we use a generalized KT-type argument as well as Monte-Carlo simulation to show that this plasma is in the screening phase, thus the free-energy cost asymptotes to constants at long distances. This result verifies Eq.~\eqref{eq:phi phi theta theta} in the main text, which is further applied to construct the effective theory for the decohered MR state away from the strong decoherence limit.

First, we note that the plasma \eqref{eq:plasma_MR_fully} is composed of one charge-$2\sqrt{m}$ Laughlin plasma and two charge-$\pm\sqrt{3}$ Ising plasmas.  
The electrons carry all three types of charges $(Q_L,Q_{I_1},Q_{I_2})=(+2\sqrt{m},+\sqrt{3},+\sqrt{3})$, and their coordinates are labeled by $\{\xi_i\}$. 
The two Ising-type plasmas contain additional particles, that carry charges $(0,-\sqrt{3},0)$ and $(0,0,-\sqrt{3})$, with coordinates $\{w^1_k\}$ and $\{w^2_l\}$, respectively. Since all three types of plasmas are in the screening phase when treated individually (for $m<35$), the combined plasma is expected to be in the screening phase.  
This can be verified using a KT-type argument. Since in the experimentally relevant region, $m$ is much smaller than $35$, the Laughlin plasma always screens. We focus on the screening property of the two Ising plasmas and omit the $Q_L$ component of the charge for simplicity. The Ising plasmas do not screen only when all particles form neutral bound trimers, which consist of one charge-$(+\sqrt{3},+\sqrt{3})$ particle, one charge-$(-\sqrt{3},0)$ particle and one charge-$(0,-\sqrt{3})$ particle. Now consider the perturbation that unbinds, without loss of generality, the charge-$(-\sqrt{3},0)$ particle from a trimer to an $O(L)$ distance. This unbinding entails a Coulomb energy cost $\Delta E\sim 3\log L$ and entropy gain $T\Delta S\sim 4\log L>\Delta E$. Consequently the combined plasma has an unbinding instability and  it is in the screening phase. 

We also numerically confirm the screening property via a Monte Carlo simulation of the combined plasma for $m=2$. 
We simulate the combined plasma on a periodic $L\times L$ square lattice with lattice constant $a=1$.
On each lattice site, there can be $n_i^{(1)}$ electrons, $n_i^{(2)}$ and $n_i^{(3)}$ the other two particles, where $n^{(s)}_i=0,1$ for $s=1,2,3$.
They carry three types of charges, labeled by $\alpha=L,I_1,I_2$.
The three types of total charges on site $i$, labeled by $q_i^{(\alpha)}$, are
\[
\begin{gathered}
q^{(L)}_i=2\sqrt{m}n^{(1)}_i\,,\\
q^{(I_1)}_i=\sqrt{3}(n^{(1)}_i-n^{(2)}_i)\,,\, q^{(I_2)}_i=\sqrt{3}(n^{(1)}_i-n^{(3)}_i)\,.
\end{gathered}
\]
The Hamiltonian of the combined plasma is
\[
\begin{aligned}
\Phi=\sum_{\alpha=L,I_1,I_2}\sum_{i, j}q^{(\alpha)}_i q^{(\alpha)}_j \Big( & \frac{1-\delta_{ij}}{2}U(\mathbf{r}_i-\mathbf{r}_j) \\
&\; -\ln y (1 - \delta_{\alpha,L}) \delta_{ij} \Big)\,, \\
\end{aligned}
\label{eq:lattice_gas_H}
\]
which contains a Coulomb interaction $U(\mathbf{r})$ and a finite fugacity $0<y<1$ for unpaired Ising charges.
The Coulomb potential $U(\mathbf{r})$ is determined by the discretized Poisson's equation
\begin{equation}
\begin{aligned}
\sum_{i=x,y}U(\mathbf{r}+a\hat e_i) + U(\mathbf{r}-a\hat e_i) - & 2U(\mathbf{r}) \\
= & -2\pi (\delta_{\mathbf{r},0}-\frac{1}{L^2})\,,
\end{aligned}
\end{equation}
The solution is:
\[
U(\mathbf{r})=\frac{1}{L^2}\sum_{|\mathbf{k}|\neq 0}\frac{\pi e^{i\mathbf{k}\cdot \mathbf{r}}}{2-\cos k_xa-\cos k_ya}\,,
\]
where $k_{x,y}=2\pi j_{x,y}/L, j_{x,y}=0,1,\cdots L-1$. In the continuum limit $1\ll |\mathbf{r}|\ll L$, $U(\mathbf{r})\approx-\log |\mathbf{r}|$, as expected.
The fugacity corresponds to the ultra-violet regularization of the continuum integral expression that we discuss in the previous subsection. 
We fix the number of particles in the numerical simulation and will neglect the Coulomb potential from the background charge since it merely results in a constant shift of $\Phi$ on a periodic lattice. 

To determine the screening property, we calculate the inverse dielectric constant $\epsilon^{-1}_0=\lim_{|k|\rightarrow 0}\epsilon^{-1}(\mathbf{k})$ of the Ising plasma $I_1$. It is finite in an insulating phase and vanishes in a screening phase. Based on the linear response theory, we have~\cite{lee1992phase} 
\[
\epsilon^{-1}(\mathbf{k})=1-\frac{2\pi}{L^2Tk^2}\langle q^{(I_1)}_{\mathbf{k}}q^{(I_1)}_{-\mathbf{k}}\rangle,
\label{eq:epiv2}
\]
where $q^{(I_1)}_k=\sum_{\text{sites }(x,y)}q^{(I_1)}_{x,y}e^{-ik_xx-ik_y y}$ is the Fourier transformed charge density of plasma $I_1$, and $T=1$ in this problem. 

\begin{figure}
    \centering
    \includegraphics[width=1.0\linewidth]{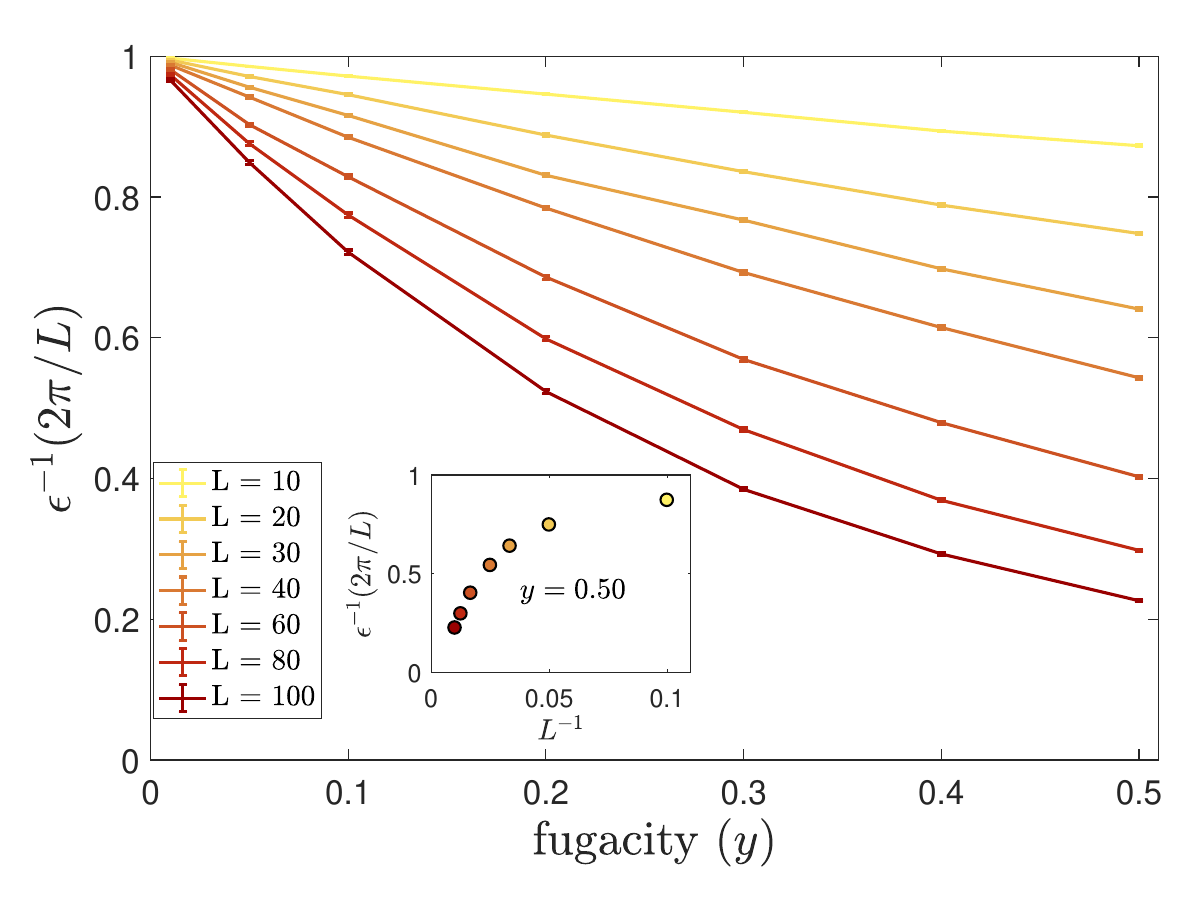}
    \caption{Inverse dielectric constant of the Ising plasma at $|\mathbf{k}|=2\pi/L$. The error bar is invisible in the plot.}
    \label{fig:MR_plasma}
\end{figure}

In the simulation, each type of particles has a fixed number $N\equiv \sum_i n_i^{(1)}=\sum_i n_i^{(2)}=\sum_i n_i^{(3)}=L^2/10$.  Initially, we generate a random configuration for each type of particles. At each step, we randomly choose (with probability $1/3$) one species of particle, a site $(r_x,r_y)$ that is occupied by that species and another site $(r_x',r_y')$ that is not. 
Then we calculate the energy difference $\Delta\Phi $ of moving the particle from $(r_x,r_y)$ to $(r'_x,r'_y)$. Then we accept or reject the update according to the standard Metropolis algorithm, that is, we move the particle from $(r_x,r_y)$ to $(r'_x,r'_y)$ with probability $p=\text{min}\{1,e^{-\Delta\Phi}\}$. For each parameter, we take average over 96 independent trajectories. In each trajectory, we discard the initial $50000 \times 3N$ steps to equilibrate the system. After equilibration, we use $50000\times 3N$ steps to evaluate the correlation function $\langle q^{(I_1)}_{\mathbf{k}}q^{(I_1)}_{\mathbf{-k}}\rangle$ in \eqnref{eq:epiv2}. 
Finally, for each system size, we average over $\epsilon^{-1}(\mathbf{k})$ for $\mathbf{k}=(2\pi/L,0)$ and $\mathbf{k}=(0,2\pi/L)$. In Fig.~\ref{fig:MR_plasma}, we show the results for the fugacity ranging from $0.01$ to $0.5$. Indeed, there is no sign of phase transition (curves for different $L$'s should exhibit crossings if there is a BKT transition, as shown in \cite{lee1992phase}), and the dielectic constant keeps decreasing with the system size, indicating that the plasma remains in the screening phase. 

\section{R\'enyi-1 correlation of the fully decohered Moore-Read states}\label{app:Renyi1_MR}

In this appendix, we show how to use plasma analogy to bound the long-distance behavior of the R\'enyi-1 correlation for the fully decohered MR states:
\begin{equation*}
\begin{aligned}
    C_e^{(1)} & (\eta,\eta') \\
    =& \frac{Z_{\text{MR}_\infty}^{-1}}{(N-1)!}\int\prod_i^{N-1} d^2z_i |\Psi(\{z_i,\eta\})\Psi(\{z_i,\eta'\})|\,,
\end{aligned}
\end{equation*}
where $\Psi(\{z_i,\eta\})$ is the MR wave function with an electron fixed at the location $\eta$. 
Plugging in the MR wave function on the plane, we have
\begin{equation}
\begin{aligned}
    |\Psi & \left(\{z_i,\eta\}\right)\Psi(\{z_i,\eta'\})| \\
    = &|\eta-\eta'|^{-m/2} e^{-\Phi_L(z_i,\sqrt{2m};\{\eta,\eta'\},\sqrt{m/2})}\\
    &\times \Big|\text{Pf}\left(\frac{1}{z_i-z_j}\right)_{z_N=\eta}\Big| \cdot \Big|\text{Pf}\left(\frac{1}{z_i-z_j}\right)_{z_N=\eta'}\Big|\,.
\end{aligned}
\label{eq:Renyi1_MR2}
\end{equation}
Here $\Phi_L(z_i,\sqrt{2m};\{\eta,\eta'\},\sqrt{m/2})$ is the energy of the Laughlin plasma with two test charges at $\eta$ and $\eta'$, both of which carry charge $\sqrt{m/2}$. The absolute value of the Pfaffian wave function does not have a direct plasma analogy on its own. 
Below we find its upper and lower bounds that do admit a plasma description, and we apply it to compute the bounds explicitly.

For the upper boundary, we use the inequality $2ab \leq a^2+b^2$ to turn the absolute value into norm squares
\[
\begin{gathered}
    2\Big|\text{Pf}\left(\frac{1}{z_i-z_j}\right)_{z_N=\eta}\Big|\cdot \Big|\text{Pf}\left(\frac{1}{z_i-z_j}\right)_{z_N=\eta'}\Big|\\
    \leq \Big|\text{Pf}\left(\frac{1}{z_i-z_j}\right)_{z_N=\eta}\Big|^2 + \Big|\text{Pf}\left(\frac{1}{z_i-z_j}\right)_{z_N=\eta'}\Big|^2\,.
\end{gathered}
\]
From Eq.~\eqref{eq:Ising_plasma}, we can interpret each term on the right-hand side as a two-component plasma with a single test charge, e.g.,
\begin{equation*}
    \Big|\text{Pf}\left(\frac{1}{z_i-z_j}\right)_{z_N=\eta}\Big|^2 = \int \prod_{k=1}^{N} d^2 w_k e^{-\Phi_I(\{z_i,\eta\},\{w_k\})}\,.
\end{equation*}
As a result, the upper bound on the R\'enyi-1 correlator has a plasma analogy as
\begin{equation*}
\begin{aligned}
    \frac{1}{2} & \frac{1}{|\eta-\eta'|^{m/2}} \frac{Z_{\text{MR}_\infty}^{-1}}{(N-1)!} \times \\
    & \int \prod_{i=1}^{N-1} d^2z_i \prod_{k=1}^N d^2 w_k e^{-\Phi_L\big(\{z_i\},\sqrt{2m};~\{\eta,\eta'\},\sqrt{m/2} \big)} \\
    & \quad (e^{-\Phi_I(\{z_i,\eta\},\{w_k\})} + e^{-\Phi_I(\{z_i,\eta'\},\{w_k\})})
\end{aligned}
\end{equation*}
Using the screening property of the plasma, we have 
\begin{equation}
    C_e^{(1)}(\eta,\eta') \leq \frac{c}{|\eta-\eta'|^{m/2}}\,, \quad (|\eta-\eta'|\gg \xi)
\end{equation}
for some positive constant $c$.

For the lower bound, we first use Eq.~\eqref{eq:Ising_plasma} to get
\begin{equation*}
\begin{aligned}
    &\Big|\text{Pf}\left(\frac{1}{z_i-z_j}\right)_{z_N=\eta}\Big|\cdot \Big|\text{Pf}\left(\frac{1}{z_i-z_j}\right)_{z_N=\eta'}\Big|\\
    = &\prod_{i<j}^{N-1}|z_i-z_j|^3 \prod_i^{N-1} |z_i-\eta|^{3/2} |z_i-\eta'|^{3/2} \\
    \times &\sqrt{\int \tilde Dw \prod_j^N|w_j-\eta|^{-3} \int \tilde Dw\prod_k^N|w_k-\eta'|^{-3}}\,,
\end{aligned}
\end{equation*}
where we introduce the short-hand notation 
\begin{equation*}
    \tilde Dw \equiv \prod_j^N d^2w_j \prod_{j<k}^N |w_j-w_k|^{3} \prod_i^{N-1} \prod_j^{N} |z_i-w_j|^{-3}\,.
\end{equation*}
\vspace{2mm}
Next, we use the Cauchy-Schwartz inequality:
\begin{equation*}
\begin{aligned}
&\int \tilde Dw \prod_j^N|w_j-\eta|^{-3} \int \tilde Dw\prod_k^N|w_k-\eta'|^{-3}\\
 \geq &\Big(\int \tilde Dw\prod_j^N|w_j-\eta|^{-3/2}|w_j-\eta'|^{-3/2} \Big)^2\,,
\end{aligned}
\end{equation*}
to obtain a plasma analogy for the lower bound
\begin{equation*}
\begin{aligned}
    &\frac{1}{|\eta-\eta'|^{m/2+3/4}} \frac{Z^{-1}}{(N-1)!}\int \prod_i^{N-1} d^2z_i\int \prod_j^N d^2 w_j \\
    &e^{-\Phi_L\big(\{z_i\},\sqrt{2m};~\{\eta,\eta'\},\sqrt{\frac{m}{2}} \big)-\Phi_I\big(\{z_i\},\{w_j\};~\{\eta,\eta'\}, \frac{\sqrt{3}}{2} \big)}
\end{aligned}
\label{eq:MR_Renyi1_lower}
\end{equation*}
where 
\begin{equation*}
\begin{gathered}
\Phi_I(z_i,w_k;\{\eta,\eta'\},\sqrt{3}/2)=\Phi_I(z_i,w_k)-\frac{3}{4}\log|\eta-\eta'|\\
-\frac{3}{2}\sum_i\log|z_i-\eta|+\frac{3}{2}\sum_k\log |w_k-\eta|,
\end{gathered}
\end{equation*}
is the energy of Ising plasma with two test charges of charge $\sqrt{3}/2$ inserted at $\eta,\eta'$. 
Using the screening property of the combined plasma, we have 
\begin{equation}
    C_e^{(1)}(\eta,\eta') \geq \frac{c'}{|\eta-\eta'|^{m/2+3/4}} \,,\quad (|\eta-\eta'|\gg \xi)\,
\end{equation}
for some positive constant $c'$.

\bibliography{ref.bib}

\end{document}